\documentclass[12pt]{article}
\pdfoutput=1

\newif{\ifdraft}

\usepackage[a4paper,vmargin=2.5cm,hmargin=2cm]{geometry}
\usepackage{lmodern}
\usepackage[T1]{fontenc}
\usepackage{amsmath,amssymb}
\usepackage[bookmarks=true,hyperfigures=true]{hyperref}
\usepackage{graphicx} 
\usepackage[font=small,labelfont=bf,width=0.85\textwidth]{caption} 
\usepackage[nosort]{cite} 
\usepackage[bulletsep]{collref} 
\usepackage{fixmath} 
\usepackage{booktabs} 
\usepackage{microtype}
\providecommand{\microtypesetup}[1]{}

\makeatletter
\renewcommand*\l@section[2]{%
  \ifnum \c@tocdepth >\z@
    \addpenalty\@secpenalty
    \addvspace{0.4em \@plus\p@}%
    \setlength\@tempdima{1.5em}%
    \begingroup
      \parindent \z@ \rightskip \@pnumwidth
      \parfillskip -\@pnumwidth
      \leavevmode \bfseries
      \advance\leftskip\@tempdima
      \hskip -\leftskip
      #1\nobreak\hfil \nobreak\hb@xt@\@pnumwidth{\hss #2}\par
    \endgroup
  \fi}
\makeatother

\allowdisplaybreaks

\graphicspath{{./figures/}}

\makeatletter \let\@keywords\@empty \let\@subject\@empty
\providecommand{\keywords}[1]{\gdef\@keywords{#1}}
\providecommand{\subject}[1]{\gdef\@subject{#1}}
\def\thetitle{\@title}
\def\theauthor{\@author}
\def\thesubject{\@subject}
\def\thedate{\@date}
\def\thekeywords{\@keywords}
\makeatother
\AtBeginDocument{%
\hypersetup{pdftitle={\thetitle}}%
\hypersetup{pdfauthor={\theauthor}}%
\hypersetup{pdfsubject={\thesubject}}%
\hypersetup{pdfkeywords={\thekeywords}}%
}

\providecommand{\hypersetup}[1]{}

\hypersetup{plainpages=false}
\hypersetup{pdfpagemode=UseNone}
\hypersetup{bookmarksnumbered=true}
\hypersetup{pdfstartview=FitH} 
\hypersetup{colorlinks=false}
\hypersetup{citebordercolor={.5 1 .5}}
\hypersetup{urlbordercolor={.5 1 1}}
\hypersetup{linkbordercolor={1 .7 .7}}

\numberwithin{equation}{section}

\let\oldbfseries=\bfseries
\let\oldmdseries=\mdseries
\let\oldnormalfont=\normalfont
\renewcommand{\bfseries}{\oldbfseries\boldmath}
\renewcommand{\mdseries}{\oldmdseries\unboldmath}
\renewcommand{\normalfont}{\oldnormalfont\unboldmath}

\newcommand{\sfrac}[2]{{\textstyle\frac{#1}{#2}}}
\newcommand{\half}{\sfrac{1}{2}}
\newcommand{\ihalf}{\sfrac{i}{2}}

\newcommand{\op}[1]{\mathcal{#1}}

\newcommand{\opec}{C} 
\newcommand{\coeff}{\mathcal{C}} 
\newcommand{\gen}[1]{\mathfrak{#1}}

\newcommand{\alg}[1]{\mathfrak{#1}}
\newcommand{\grp}[1]{\mathrm{#1}}
\newcommand{\superN}{\mathcal{N}}
\newcommand{\Nc}{N_{\mathrm{c}}}

\newcommand{\order}[1]{\mathcal{O}(#1)}

\DeclareMathOperator{\tr}{\mathrm{tr}}
\DeclareMathOperator{\Li}{\mathrm{Li}}
\newcommand{\trb}[1]{\tr\sbrk{#1}}

\newcommand{\cR}{\ensuremath{\mathcal{R}}}

\newcommand{\cX}{\ensuremath{\mathcal{X}}}
\newcommand{\cY}{\ensuremath{\mathcal{Y}}}
\newcommand{\cZ}{\ensuremath{\mathcal{Z}}}

\newcommand{\Complex}{\mathbb{C}}

\newcommand{\dd}{d}

\newcommand{\eps}{\varepsilon}

\newcommand{\bfm}[1]{\boldsymbol{#1}}

\newcommand{\supup}[1]{^{\mathrm{#1}}}

\newcommand{\nn}{\nonumber}

\newcommand{\brk}[1]{(#1)}
\newcommand{\lrbrk}[1]{\left(#1\right)}
\newcommand{\bigbrk}[1]{\bigl(#1\bigr)}
\newcommand{\biggbrk}[1]{\biggl(#1\biggr)}
\newcommand{\Bigbrk}[1]{\Bigl(#1\Bigr)}
\newcommand{\Biggbrk}[1]{\Biggl(#1\Biggr)}
\newcommand{\sbrk}[1]{[#1]}

\newcommand{\biggsbrk}[1]{\biggl[#1\biggr]}
\newcommand{\Bigsbrk}[1]{\Bigl[#1\Bigr]}
\newcommand{\Biggsbrk}[1]{\Biggl[#1\Biggr]}
\newcommand{\brc}[1]{\{#1\}}

\newcommand{\vev}[1]{\langle#1\rangle}

\newcommand{\bigvev}[1]{\bigl\langle#1\bigr\rangle}
\newcommand{\comm}[2]{[#1,#2]}

\newcommand{\abs}[1]{|#1|}

\newcommand{\floor}[1]{\lfloor#1\rfloor}

\makeatletter
\def\mr@ignsp#1 {\ifx\:#1\@empty\else #1\expandafter\mr@ignsp\fi}%
\newcommand{\multiref}[1]{\begingroup
\xdef\mr@no@sparg{\expandafter\mr@ignsp#1 \: }%
\def\mr@comma{}%
\@for\mr@refs:=\mr@no@sparg\do{\mr@comma\def\mr@comma{,\,}\ref{\mr@refs}}%
\endgroup}
\makeatother

\newcommand{\hypref}[2]{\ifx\href\asklfhas #2\else\href{#1}{#2}\fi}
\newcommand{\secref}[1]{Section~\multiref{#1}}
\newcommand{\appref}[1]{Appendix~\multiref{#1}}
\newcommand{\tabref}[1]{Table~\multiref{#1}}
\newcommand{\figref}[1]{Figure~\multiref{#1}}
\renewcommand{\eqref}[1]{(\multiref{#1})}

\makeatletter
\newlength{\apb@width}
\newcommand{\autoparbox}[2][c]{\settowidth{\apb@width}{#2}\parbox[#1]{\apb@width}{#2}}
\newcommand{\includegraphicsbox}[2][]{\autoparbox{\includegraphics[#1]{#2}}}
\makeatother

\ifdraft
\usepackage{xcolor}
\newcommand{\remark}[1]{{\renewcommand{\bfdefault}{b}{\color[RGB]{0,150,0}{\textbf{#1}}}}}
\fi
\providecommand{\remark}[1]{\ignorespaces}

\title{Four-Point Functions with a Twist}

\author{Till Bargheer}

\begin{document}

\pdfbookmark[1]{Title Page}{title}

\thispagestyle{empty}

\begin{flushright}
\texttt{DESY 16-248}
\end{flushright}

\vfill
\vfill

\begin{center}

\begingroup\Large\bfseries\thetitle\par\endgroup

\vfill

\begingroup\scshape
Till Bargheer\par
\endgroup

\medskip

\begingroup\itshape
DESY Theory Group, DESY Hamburg,\\
Notkestra\ss e 85, D-22603 Hamburg, Germany
\endgroup

\medskip

{\ttfamily
\href{mailto:till.bargheer@desy.de}{till.bargheer@desy.de}
}

\vfill

\textbf{Abstract}

\bigskip

\begin{minipage}{12.5cm}
We study the OPE of correlation functions of local operators in planar $\superN=4$
super Yang--Mills theory. The considered operators have an explicit spacetime
dependence that is defined by twisting the translation generators with
certain R-symmetry generators. We restrict to operators that carry a
small number of excitations above the twisted BMN vacuum. The OPE limit of
the four-point correlator is dominated by internal states with few
magnons on top of the vacuum. The twisting directly couples all
spacetime dependence of the correlator to these magnons.
We analyze the OPE in detail, and
single out the extremal states that have to cancel all double-trace
contributions.
\end{minipage}

\end{center}

\vfill
\vfill
\vfill

\setcounter{page}{0}


\newpage

\setcounter{tocdepth}{1}
\hrule height 0.4pt
\pdfbookmark[1]{\contentsname}{contents}
\microtypesetup{protrusion=false}
\tableofcontents
\microtypesetup{protrusion=true}
\vspace{0.6cm}
\hrule height 0.4pt
\vspace{1cm}


\section{Introduction}
\label{sec:intro}

Solving planar $\superN=4$ super Yang--Mills theory (sYM) at any value of
its coupling constant continues to be an important goal in
mathematical physics. This most symmetric four-dimensional gauge
theory serves as a key towards understanding general properties of
interacting gauge theory, AdS/CFT duality, and quantum gravity.
In probing the theory at ever deeper levels, we have witnessed
extraordinary progress due to the emergence of
integrability~\cite{Beisert:2010jr}. By now, the spectrum of
single-trace operators is virtually solved, for any value of the
coupling constant~\cite{Gromov:2013pga,Gromov:2014caa}.
In the past few years, also three-point functions of local operators,
as well as scattering amplitudes have come into focus. For both types
of observables, essential proposals for finite-coupling descriptions
have been made~\cite{Basso:2013vsa,Basso:2013aha,Basso:2015zoa}, and
highly non-trivial implications could be derived from these
proposals~\cite{Basso:2014pla,Basso:2015uxa}.
At this point, it appears
not unlikely that all correlation functions will eventually be computable at
any value of the coupling, at least in the planar limit.

In this paper, we focus on four-point correlation functions of local
gauge-invariant operators. While
two- and three-point functions in principle determine all higher-point
functions via the conformal operator product expansion, actually
resumming the expansion to recover explicit higher-point functions is
very difficult in practice. Conversely, four-point (and higher-point)
functions contain a wealth of information on structure constants and
scaling dimensions, and are thus very interesting physical objects in
their own right. It therefore remains desirable to devise efficient
methods for computing higher-point correlators directly. Here, our
primary goal will be to find the right language and variables for
making use of the integrable structure for general correlation
functions.

A key point of the integrability-based solution to the spectral
problem is the organization of local operators in terms of excitations
above a ``vacuum'' BPS operator $\trb{Z^J}$, where
$Z$ is a complex scalar. The spectrum of anomalous scaling dimensions
is then encoded in the eigenstates and
eigenvalues of the dilatation generator, which are governed by the Beisert
scattering matrix among the set of excitations (scalars, fermions, and
covariant derivatives) above the $\trb{Z^J}$ vacuum~\cite{Beisert:2005tm}. In particular, the
scattering matrix is, up to its overall phase, completely constrained
by the $\alg{psu}(2|2)^2$ subalgebra of the superconformal
symmetry that preserves the vacuum.

For higher-point functions, it is again desirable to consider operators that
preserve as much symmetry as possible. The familiar BPS operators $\trb{Z^J}$
suggest themselves. But when considering more than two insertions of
such operators, most of the symmetry is broken, because the
$\alg{psu}(2|2)^2$ symmetries of the individual insertions are not aligned.
The situation can be improved by
considering slightly generalized BPS operators, which are related to
$\trb{Z^J}$ by an internal rotation. Namely, consider
$\trb{\phi^J}$, where $\phi=X\cdot\Phi$ and $X\in\Complex^6$, $X^2=0$
specifies a complex, lightlike direction in the internal space of
scalar fields $\Phi^I$, $I=1,\dots,6$. By judiciously
choosing the vector $X$ as a function of the spacetime coordinate, a
substantial part of the superconformal symmetry can be preserved. This
has been noted before~\cite{deMedeiros:2001kx,Drukker:2009sf}.
Namely, after choosing a vector $X$ at the spacetime origin $x=0$,
define $X(x)$ throughout spacetime via a twisted translation
generator $\check{\alg{P}}\sim\alg{P}-\alg{R}$ that is a combination of the conventional spacetime
translation $\alg{P}$ and an internal rotation $\alg{R}$.
This twisted translation is associated to
a whole \emph{twisted conformal symmetry algebra}
$\widehat{\alg{so}}(2,4)$, which is a diagonal combination of the original conformal
algebra $\alg{so}(2,4)$ and the internal $\alg{so}(6)$ R-symmetry.
The twisted fields were first considered in the context of topological
twistings of $\superN=4$ super Yang--Mills theory~\cite{deMedeiros:2001kx}.
Later, correlation functions of single-trace operators of a single twisted
complex scalar $\phi$ were considered~\cite{Drukker:2009sf}, and found
to (i) have trivial spacetime dependence, and (ii) be protected from
quantum corrections. These properties make such operators good candidates
for an integrability-based description in terms of excitations on top
of vacuum operators $\trb{\phi\dots\phi}$.
In fact, the hexagon form factor proposal for three-point
functions~\cite{Basso:2015zoa} relies on an excitation picture around
vacuum operators of this type.

In this paper, we consider four-point functions of twisted
single-trace operators with few excitations at the one-loop level. For the time being, the
analysis is restricted to zero-momentum excitations. While this means that
the four operators are still half-BPS, their four-point functions are
not protected. We extract the one-loop
correlation functions of such operators from~\cite{Drukker:2008pi},
and analyze their OPE decomposition. In
the double-coincidence limit $\abs{x_{12}}, \abs{x_{34}}\ll\abs{x_{23}}$, the one-loop
correlator displays a logarithmic singularity, which is generated by the
anomalous dimensions of the internal states in the OPE. We show how
this leading term is captured by states with a minimal number of
excitations on top of the twisted vacuum. For ``extremal'' OPE
contributions (which stem from internal states with the maximal
R-charge compatible with the external operators), it is known that
there is a $1/\Nc^2$ mixing with double-trace operators. We
find that the contribution of double-trace operators can be neglected
by simply projecting out specific extremal states in the OPE.

\medskip
\noindent
\textbf{Note:} While this work was being completed, I learned of the very interesting
parallel paper~\cite{Basso:2017khq}, which also discusses the OPE of four BPS
operators from an integrability perspective. While this work considers
operators of any charge,~\cite{Basso:2017khq} focuses on operators
with large charges and the relation to the hexagon form factor approach.

\section{Twisted States}

We start by reviewing essential parts of~\cite{deMedeiros:2001kx},
\cite{Drukker:2009sf}, and \cite{Beisert:2002tn}.
Consider a general complex linear combination
$X\cdot\Phi=X_I\Phi^I$ of the six real scalar fields $\Phi^I$. The
propagator for two such fields equals
(neglecting color factors)
\begin{equation}
\bigvev{X_1\cdot\Phi(x_1)\,X_2\cdot\Phi(x_2)}
=
\frac{X_1\cdot X_2}{(2\pi)^2\,x_{12}^2}
\,.
\end{equation}
Correlation functions of single-trace operators are difficult to
compute in general, and have a complicated dependence on the spacetime
coordinates as well as the coupling constant,
unless they obey extra relations due to supersymmetry.
A prominent case are half-BPS chiral primary operators
$\trb{\brk{X\cdot\Phi}^J}$ with $X^2=0$. Every such operator
preserves $24$ supersymmetries. Therefore any two such operators share
$16$ supersymmetries, and any three such operators share $8$
supersymmetries. This is the reason why two-point and three-point
functions of such half-BPS operators are protected from quantum
corrections. The same is generically not true for correlation
functions of four or more such operators, which generically do not
share any supersymmetry. However, making the
judicious, spacetime-dependent choice
\begin{gather}
X=X^\phi(x^\mu)
\equiv
\sqrt{2}\bigbrk{ix^\mu,\half(1+x^2),\ihalf(1-x^2)}
\,,
\label{eq:Xphi}
\\
\phi(x)\equiv
X^\phi(x)\cdot\Phi(x)
\label{eq:phi}
\end{gather}
has two important consequences. Firstly,
$X^\phi(x_1)\cdot{}X^\phi(x_2)=x_{12}^2$, therefore the free propagator of two
fields $\phi$ becomes constant,
\begin{equation}
\vev{\phi(x_1)\,\phi(x_2)}=\frac{1}{(2\pi)^2}\,.
\label{eq:phiprop}
\end{equation}
Hence, in the free field theory, correlation functions of any number
of operators
\begin{equation}
\op{Q}^J(x)\equiv\trb{\phi(x)^J}
\label{eq:vacop}
\end{equation}
are completely spacetime-independent and just evaluate to constants.
Secondly, all correlation functions
\begin{equation}
\bigvev{\op{Q}^{J_1}(x_1)\dots\op{Q}^{J_n}(x_n)}
\end{equation}
are protected from quantum corrections, for any number $n$ of operator
insertions. The reason is that all operators $\op{Q}^J(x)$ share two
universal supercharges $\alg{Q}^\pm$, and the perturbative action of
$\superN=4$ sYM is
$\alg{Q}^\pm$-exact~\cite{deMedeiros:2001kx,Drukker:2009sf}.%
\footnote{See also~\cite{Liendo:2016ymz} for an OPE-based argument
that does not refer to the Lagrangian.}
We will refer to the
operators~\eqref{eq:vacop} as \emph{vacuum operators}.

The field $\phi(x)$ can be defined  via a translation that is
twisted with an appropriate R-symmetry generator. In fact, the
field~\eqref{eq:phi} was first considered in the context of
topological twists of $\superN=4$ sYM~\cite{deMedeiros:2001kx}.
The twisted translation is associated with
a whole \emph{twisted conformal symmetry algebra}
$\widehat{\alg{so}}(2,4)$, which is a diagonal combination of the original conformal
algebra $\alg{so}(2,4)$ and the internal $\alg{so}(6)$ R-symmetry.
Denoting the generators of the original conformal symmetry by
$\brc{\alg{P}_\mu, \alg{K}_\mu, \alg{L}_{\mu\nu}, \alg{D}}$, the twisted
conformal symmetry generators read%
\footnote{A similar twisting was used in~\cite{Beem:2013sza} to
construct a chiral algebra of operators living in a two-dimensional
hyperplane of the four-dimensional Minkowski spacetime. The
construction of~\cite{Beem:2013sza} involves passing to the cohomology
of a suitable supercharge, upon which all operators transform
trivially under a twisted conformal algebra. On the contrary, here
we study fully four-dimensional operators that transform non-trivially
under the twisted conformal symmetry.}
\begin{alignat}{2}
\alg{\hat{P}}_\mu&=\alg{P}_\mu+\alg{R}_{\mu-}\,,\qquad&
\alg{\hat{L}}_{\mu\nu}&=\alg{L}_{\mu\nu}+\alg{R}_{\mu\nu}\,,
\nn\\
\alg{\hat{K}}_\mu&=\alg{K}_\mu+\alg{R}_{\mu+}\,,&
\alg{\hat{D}}&=\alg{D}+\alg{R}\,.
\label{eq:twistalg}
\end{alignat}
Here, the R-symmetry algebra has been split into an $\alg{so}(4)$ with
generators $\alg{R}_{\mu\nu}$ and a remainder with generators
$\alg{R}_{\mu\pm}\equiv -i\alg{R}_{\mu5}\pm\alg{R}_{\mu6}$, and
$\alg{R}\equiv -i\alg{R}_{56}$. Starting with
$\phi(0)=\brk{\Phi^5+i\Phi^6}/\sqrt{2}\equiv{}Z$ at the origin,
$\phi(x)$ throughout spacetime is obtained by
\begin{equation}
\phi(x)\equiv
\exp\brk{x^\mu\brk{\alg{P}_\mu-\alg{R}_{\mu-}}}\cdot Z(0)
=\exp\brk{-x^\mu\alg{R}_{\mu-}}\cdot Z(x)
\,.
\label{eq:phibytrans}
\end{equation}
The splitting is chosen such that
$\alg{R}_{\mu+}$ annihilates $\phi(0)=Z$ at the origin, and
$\alg{R}$ measures the R-charge along the direction of $Z$,
\begin{equation}
\alg{R}\cdot Z=+Z\,,
\qquad
\alg{R}\cdot\bar Z=-\bar Z\,,
\qquad
\comm{\alg{R}}{\alg{R}_{\mu\pm}}=\pm\alg{R}_{\mu\pm}\,.
\end{equation}
Note that $\gen{R}_{\mu-}$ appears with opposite signs
in~\eqref{eq:twistalg} and~\eqref{eq:phibytrans}. This is natural: The
exponential in~\eqref{eq:phibytrans} ``undoes'' the twisting defined
by~\eqref{eq:twistalg}, such that the twisted translation generator
$\gen{\hat{P}}_\mu$ simply acts as
\begin{equation}
\gen{\hat{P}}_\mu\phi=\partial_\mu\phi\,,
\end{equation}
where $\partial_\mu$ only acts on the fields $\Phi^I$, and not on the
explicit spacetime coordinates $x$.
Together with $\phi$, the remainder of the six scalars $\Phi^I$
naturally organizes into a multiplet
$\brc{\phi,V_\mu,B}$ of the twisted conformal symmetry, with
\begin{alignat}{2}
V_\mu
&\equiv X^{V_\mu}\cdot\Phi
\,,&\qquad
X^{V_\mu}
&\equiv\sfrac{1}{\sqrt{2}}(\partial_\mu X^\phi)
=\brk{ie_\mu,x_\mu,-ix_\mu}
\,,
\\
B
&\equiv X^B\cdot\Phi
\,,&\qquad
X^B
&\equiv\sfrac{1}{4\sqrt{2}}(\partial^2X^\phi)
=\brk{0,0,0,0,1,-i}
\,.
\end{alignat}
Just as $\phi(x)$, the fields $V_\mu(x)$ and $B(x)$ throughout
spacetime are obtained by the same twisted
translation~\eqref{eq:phibytrans} applied to $V_\mu(0)$ and $B(0)$ at
the origin. The latter reduce to
\begin{equation}
V_\mu(0)=i\Phi_\mu
\,,\qquad
B(0)=\Phi_5-i\Phi_6=\sqrt{2}\,\bar Z
\,.
\end{equation}
The twisted translation is defined such that, when acting on
the twisted fields $\phi$, $V_\mu$, and $B$,
the generators of the twisted conformal algebra $\widehat{\alg{so}}(2,4)$
take the standard realization
\begin{align}
\alg{\hat{P}}_\mu&=\partial_\mu\,,\nn\\
\alg{\hat{L}}_{\mu\nu}&=(x_\mu\partial_\nu-x_\nu\partial_\mu)+\hat\Sigma_{\mu\nu}\,,\nn\\
\alg{\hat{D}}&=x^\mu\partial_\mu+\hat\Delta\,,\nn\\
\alg{\hat{K}}_\mu&=2x_\mu x\cdot\partial-x^2\partial_\mu-2x_\mu\hat\Delta+2x^\nu\Sigma_{\mu\nu}+\hat\kappa_\mu\,,
\end{align}
where the derivatives only act on the fields $\Phi^I$, and not on the
explicit coordinates $x^\mu$ in the definition of the fields $\phi$,
$V_\mu$, and $B$. The generators $\hat\Delta$, $\hat\Sigma_{\mu\nu}$
and $\hat\kappa_\mu$ specify the transformation
properties of the fields at the origin,
\begin{alignat}{3}
\comm{\hat\Delta}{\phi}&=0\,,&\qquad
\comm{\hat\Sigma_{\mu\nu}}{\phi}&=0\,,&\qquad
\comm{\hat\kappa_\mu}{\phi}&=0\,,\nn\\
\comm{\hat\Delta}{V_\rho}&=V_\rho\,,&\qquad
\comm{\hat\Sigma_{\mu\nu}}{V_\rho}&=\eta_{\rho\mu}V_\nu-\eta_{\rho\nu}V_\mu\,,&\qquad
\comm{\hat\kappa_\mu}{V_\rho}&=-\eta_{\mu\rho}\sqrt{2}\,\phi\,,\nn\\
\comm{\hat\Delta}{B}&=2B\,,&\qquad
\comm{\hat\Sigma_{\mu\nu}}{B}&=0\,,&\qquad
\comm{\hat\kappa_\mu}{B}&=-2V_\mu\,.
\label{eq:kappa}
\end{alignat}
That is $\phi$ is a scalar with dimension zero, $V_\mu$ is a vector
with dimension one, and $B$ is a scalar with dimension two under the
twisted conformal symmetry. More generally, by construction, the
twisted scaling dimension of any operator is the sum of the untwisted
scaling dimenion and the R-charge in the $Z$ direction. What is
nonstandard is that the special conformal generator
$\alg{\hat{K}}$ acts non-trivially at the origin, even though the
fields $V_\rho$ and $B$ are not conformal descendants.%
\footnote{In the classification of representations of the conformal
algebra by Mack and Salam~\cite{Mack:1969rr}, the
representation~\eqref{eq:kappa} belongs to class Ib ($\kappa_\mu\neq0$
but nilpotent).}
Due to the non-trivial action of $\hat\kappa$, the representation of
the little group (with generators $\hat\Sigma_{\mu\nu}$, $\hat\Delta$,
and $\hat\kappa_\mu$) is not further reducible, even though $\phi$,
$V_\mu$, and $B$ have different scaling dimensions. This affects the
conformal Ward identities for correlation functions, and hence
correlators of twisted fields generically do not have the standard
form (that follows from the Ward identities for $\hat\kappa_\mu=0$).
In particular, operators with different twisted dimensions may have
non-vanishing two-point functions.

The twisting is explained in detail
in~\cite{deMedeiros:2001kx,Drukker:2009sf}. Here, we only note that
under the twisted conformal symmetry,
the fermion fields
organize into vectors $\psi_\mu^{(1)}$, $\tilde\psi_\mu^{(1)}$,
two-form fields $\chi^{\pm,(2)}_{\mu\nu}$, and scalars $\eta^{(2)}$,
$\tilde\eta^{(2)}$, where the superscript numbers denote the twisted
scaling dimensions. The gauge fields are R-symmetry singlets and thus
not affected by the twisting. The twisted matter fields
$\brc{\phi,V_\mu,B,\psi_\mu,\tilde{\psi_\mu},\chi^\pm_{\mu\nu},\eta,\tilde\eta}$
transform covariantly under fifteen combinations $\alg{Q}_\mu$,
${\alg{S}}_\mu$, $\alg{Q}_{\mu\nu}$, and $\alg{Q}_{\mathrm{D}}$ of the
fermionic generators of $\alg{psu}(2,2|4)$. Commuting the latter with
the singlets $\alg{Q}^\pm$ that preserve the vacuum field $\phi$
yields the twisted conformal algebra,
\begin{alignat}{2}
\comm{\alg{Q}^\pm}{\alg{Q}_\mu}&=\hat{\alg{P}}_\mu\,,
\qquad &
\comm{\alg{Q}^\pm}{\alg{Q}_{\mu\nu}}&=\hat{\alg{L}}_{\mu\nu}\,,
\nn\\
\comm{\alg{Q}^\pm}{\alg{S}_\mu}&=\hat{\alg{K}}_\mu\,,
&
\comm{\alg{Q}^\pm}{\alg{Q}_{\mathrm{D}}}&=\hat{\alg{D}}\,.
\end{alignat}
The fermionic generators $\brc{\alg{Q}_\mu,\alg{S}_\mu,\alg{Q}_{\mu\nu},\alg{Q}_{\mathrm{D}}}$
and their bosonic counterparts
$\brc{\hat{\alg{P}}_\mu,\hat{\alg{K}}_\mu,\hat{\alg{L}}_{\mu\nu},\hat{\alg{D}}}$
combine into the strange superalgebra $\alg{q}(3)$~\cite{Kac:1977em}.

Besides the vacuum operators $\op{Q}^J$, we will also consider
operators with zero-momentum excitations (R-symmetry descendants of
the vacuum)
\begin{equation}
\op{Q}^J_\mu=\trb{V_\mu\phi^{J-1}}
\,,\qquad
\op{Q}^J_{(\mu\nu)}=\sum_{j=0}^{J-2}\trb{V_{(\mu}\phi^jV_{\nu)}\phi^{J-2-j}}
\,,\qquad
\dots
\,,
\label{eq:zeromomstates}
\end{equation}
and two-magnon operators with non-zero momenta
\begin{align}
\op{O}^J_p&=
\frac{1}{\sqrt{J+3}}\biggsbrk{
  \frac{1}{2}\sum_{j=0}^J\cos(p(j+3/2))\trb{V_\mu\phi^jV^\mu\phi^{J-j}}
  +\sqrt{2}\cos(p/2)\trb{B\phi^{J+1}}
}\,,\nn\\
\op{O}^{J,VV}_{p,(\mu\nu)}&=
\frac{1}{\sqrt{J+3}}\sum_{j=0}^{J+2}\cos(p(j+1/2))\trb{V_{(\mu}\phi^jV_{\nu)}\phi^{J+2-j}}
\,,\nn\\
\op{O}^{J,\partial\partial}_{p}&=
\frac{1}{\sqrt{J+3}}
\frac{1}{4}\sum_{j=0}^{J+2}\cos(p(j+1/2))\trb{\phi_{,\mu}\phi^j\phi^{,\mu}\phi^{J+2-j}}
+\dots
\,,\nn\\
\op{O}^{J,\partial\partial}_{p,(\mu\nu)}&=
\frac{1}{\sqrt{J+3}}\biggsbrk{
  \frac{1}{2}\sum_{j=0}^{J}\cos(p(j+3/2))\trb{\phi_{,(\mu}\phi^j\phi_{,\nu)}\phi^{J-j}}
  +\frac{1}{2}\cos(p/2)\trb{\phi_{,(\mu\nu)}\phi^{J+1}}
}\,,\nn\\
\op{O}^{J,V\partial,1}_{p,\mu\nu}&=
\frac{-i}{\sqrt{J+3}}\biggsbrk{
  \frac{1}{\sqrt{2}}\sum_{j=0}^{J}\cos(p(j+3/2))\trb{V_\mu\phi^j\phi_{,\nu}\phi^{J-j}}
  +\sqrt{2}\cos(p/2)\trb{V_{\mu,\nu}\phi^{J+1}}
  +\dots
}\,,\nn\\
\op{O}^{J,V\partial,2}_{p,\mu\nu}&=
\frac{-i}{\sqrt{J+3}}
\frac{1}{\sqrt{2}}\sum_{j=0}^{J+2}\cos(p(j+1/2))\trb{V_\mu\phi^j\phi_{,\nu}\phi^{J+2-j}}
+\dots
\,.
\label{eq:2magops}
\end{align}
Here, $\phi_{,\mu}=X^\phi\cdot D_\mu\Phi$ and $V_{\mu,\nu}=X^{V_\mu}\cdot
D_\nu\Phi$ are the twisted combinations of covariant
derivative fields $D_\mu\Phi^I$. These two-magnon states are simply twisted
cousins of the two-magnon states listed in Appendix~B
of~\cite{Beisert:2002tn}. They all belong to a common superconformal
multiplet whose primary is the operator $\op{O}^J_p$, where $J$ labels
the charge under the (untwisted) operator $\gen{R}$ measured at the
origin $x=0$.%
\footnote{The full supermultiplet contains further states, which
however will not be relevant, since they are either fermionic or
antisymmetric, and thus do not contribute to OPE's of operators
$\op{Q}^J$, $\op{Q}^J_\mu$ considered below.}
The ellipses in~\eqref{eq:2magops} stand for terms with fermion fields and non-linear
corrections which do not contribute to the free OPE coefficients.
In the operators~\eqref{eq:2magops}, the two magnons
carry opposite momenta~$p$ and~$-p$, with
\begin{equation}
p\in\frac{2\pi n}{J+3}
\,,\qquad
1\leq n\leq\floor{\sfrac{J+2}{2}}
\,.
\label{eq:momcond}
\end{equation}
Since the twist amounts to a
(position-dependent) R-symmetry rotation, and the one-loop dilatation
operator commutes with the R-symmetry, all eigenstates remain
eigenstates, with unmodified eigenvalues, and all states with
different eigenvalues remain orthogonal.%
\footnote{For states with identical anomalous dimension, orthogonality
may not be preserved by the twisting. In fact, states with scalar
zero-momentum excitations $V_\mu$ or $B$, which we will consider
below, are not orthogonal to the vacuum operators $\op{Q}$, even
though they are orthogonal before twisting.}
In other words, the anomalous
dimension is unaffected by the twisting. The two-magnon states
therefore have twisted dimensions
\begin{equation}
\hat\Delta(p)=2+\lambda\Delta_1(p)+\order{\lambda^2}
\,,\qquad
\Delta_1(p)=\frac{1}{\pi^2}\sin^2(p/2)
\,,
\label{eq:anomdim}
\end{equation}
with the familiar value $\Delta_1$ of the anomalous dimension.

While the operator $\op{O}^J_p(x=0)$ carries (untwisted)
$\gen{R}$-charge $J$, the charges of the superdescendant operators
$\op{O}^{J,..}_{p,..}$ are shifted.
Besides the charge $J$, we will sometimes refer to the ``length''
$L$, which equals the number of fields within a single-trace operator.
The following table summarizes the $\gen{R}$-charges (at $x=0$) and
lengths of the various operators:
\begin{equation}
\begin{tabular}{rccccccccc}
\toprule
Operator: & $\op{O}^J_p$ & $\op{O}^{J,VV}_{p,(\mu\nu)}$ &
                                                          $\op{O}^{J,\partial\partial}_{p}$
  & $\op{O}^{J,\partial\partial}_{p,(\mu\nu)}$ &
                                                 $\op{O}^{J,V\partial,1}_{p,\mu\nu}$
  & $\op{O}^{J,V\partial,2}_{p,\mu\nu}$ & $\op{Q}^J$ & $\op{Q}^J_\mu$ & $\op{Q}^J_{(\mu\nu)}$ \\
\midrule
$\gen{R}$-charge: & $J$ & $J+2$ & $J+4$ & $J+2$ & $J+1$ & $J+3$ & $J$
                                                     & $J-1$ & $J-2$ \\
Length $L$: & $J+2$ & $J+4$ & $J+4$ & $J+2$ & $J+2$ & $J+4$ & $J$ & $J$ & $J$ \\
\bottomrule
\end{tabular}
\end{equation}
%

\section{Two-Point and Three-Point Correlators}
\label{sec:two-point-three}

In the following, we will compute two-point and three-point correlators of vacuum
operators, operators with zero-momentum excitations, and two-magnon
operators, at leading order in the Yang--Mills coupling
$g_{\mathrm{YM}}$.
The three-point correlators yield the structure constants that
will later be needed for the OPE analysis of four-point correlators.
Here and in everything that follows, we omit all gauge group factors
as well as all factors of $1/(2\pi)^2$ from propagators~\eqref{eq:phiprop}.
Keeping this in mind, the classical contractions for the twisted fields are:
\begin{align}
\vev{\phi(x_1)\phi(x_2)}&=1\,,
&
\vev{V_\mu(x_1)V_\nu(x_2)}&=-\frac{\eta_{\mu\nu}}{x_{12}^2}\,,
&
\vev{B(x_1)B(x_2)}&=0\,,
\nn\\
\vev{\phi(x_1)V_\mu(x_2)}&=-\sqrt{2}\,\frac{x_{12,\mu}}{x_{12}^2}\,,
&
\vev{\phi(x_1)B(x_2)}&=\sqrt{2}\,\frac{1}{x_{12}^2}\,,
&
\vev{V_\mu(x_1)B(x_2)}&=0\,.
\label{eq:contractions}
\end{align}
The contractions for twisted fields with twisted versions of
derivative fields similarly follow by adding the twisting factor
$X_1\cdot X_2$ to the untwisted propagators, for instance
\begin{equation}
\bigvev{(X_1\cdot\Phi(x_1))\,(X_2\cdot D_\mu\Phi(x_2))}
=-2\frac{x_{12,\mu}}{x_{12}^4}(X_1\cdot X_2)\,.
\end{equation}
%

\paragraph{Two-Point Functions.}

For non-zero momenta, the two-magnon states~\eqref{eq:2magops} are
annihilated by the special conformal generator
$\hat\kappa_\mu$~\eqref{eq:kappa}. They are thus proper conformal
primaries of the
twisted conformal algebra. Hence their two-point
functions take the standard form as dictated by conformal symmetry:%
\footnote{The two-point functions for the operators
$\op{O}^{J,V\partial,1}$ and $\op{O}^{J,V\partial,2}$ get non-trivial
contributions from fermion terms not displayed in~\eqref{eq:2magops}.
For $\op{O}^{J,V\partial,1}_p$, reducing to the bosonic terms gives
$\bigbrk{1/2+\brk{1+\cos(p)}/\brk{J+3}}$ times the full two-point function. For
$\op{O}^{J,V\partial,2}$, reducing to the bosonic terms gives $1/2$ of
the full two-point function.}
\begin{align}
\bigvev{\op{O}^J_{p,1}\,\op{O}^J_{q,2}}
=
\bigvev{\op{O}^{J,\partial\partial}_{p,1}\,\op{O}^{J,\partial\partial}_{q,2}}
&=
\frac{\delta_{p,q}}{x_{12}^4}
\nn\\
\bigvev{\op{O}^{J,VV}_{p,(\mu\nu),1}\,\op{O}^{J,VV}_{q,(\rho\sigma),2}}
=
\bigvev{\op{O}^{J,\partial\partial}_{p,(\mu\nu),1}\,\op{O}^{J,\partial\partial}_{q,(\rho\sigma),2}}
&=
\frac{\delta_{p,q}}{x_{12}^4}
I_{12,\rho(\mu}I_{12,\nu)\sigma}
\nn\\
\bigvev{\op{O}^{J,V\partial,1}_{p,\mu\nu,1}\,\op{O}^{J,V\partial,1}_{q,\rho\sigma,2}}
=
\bigvev{\op{O}^{J,V\partial,2}_{p,\mu\nu,1}\,\op{O}^{J,V\partial,2}_{q,\rho\sigma,2}}
&=
\frac{\delta_{p,q}}{x_{12}^4}I_{12,\mu\rho}I_{12,\nu\sigma}
\label{eq:twopts}
\end{align}
Here,
\begin{equation}
I_{ij,\alpha\beta}\equiv I_{\alpha\beta}(x_{ij})
\,,\qquad
I_{\alpha\beta}(x)=\eta_{\alpha\beta}-\frac{2x_\alpha x_\beta}{x^2}
\end{equation}
is the inversion tensor, and parentheses denote
traceless symmetrization with unit weight:
\begin{equation}
x_{(\mu\nu)}
\equiv
\half x_{\mu\nu}+\half x_{\nu\mu}-\sfrac{1}{4}\eta_{\mu\nu}x^{\rho}_{\rho}
\,.
\end{equation}
All other two-point functions among two-magnon states with non-zero
momenta vanish.

\paragraph{Two Vacua.}

By explicit computation, one finds that the
three-point functions of two-magnon states (with non-zero
momentum) and two vacuum states of the same weight are:
\begin{align}
\bigvev{\op{O}^{J}_{p,1}\,\op{Q}^{k}_2\,\op{Q}^{k}_3}
=
\bigvev{\op{O}^{J,\partial\partial}_{p,1}\,\op{Q}^{k}_2\,\op{Q}^{k}_3}
&=
\coeff^J_pk^2\frac{x_{23}^{2}}{x_{12}^2x_{13}^2}\,,
\nn\\
\bigvev{\op{O}^{J,VV}_{p,(\mu\nu),1}\,\op{Q}^{k}_2\,\op{Q}^{k}_3}
=
\bigvev{\op{O}^{J,\partial\partial}_{p,(\mu\nu),1}\,\op{Q}^{k}_2\,\op{Q}^{k}_3}
&=
\coeff^J_p2k^2\,Y_{(\mu,12,13}Y_{\nu),12,13}\,,
\nn\\
\bigvev{\op{O}^{J,V\partial,1}_{p,\mu\nu,1}\,\op{Q}^{k}_2\,\op{Q}^{k}_3}
=
\bigvev{\op{O}^{J,V\partial,2}_{p,\mu\nu,1}\,\op{Q}^{k}_2\,\op{Q}^{k}_3}
&=
\coeff^J_p2ik^2\,Y_{\mu,12,13}Y_{\nu,12,13}\,,
\end{align}
with
\begin{equation}
Y_\alpha(x,y)=\frac{x^\alpha}{x^2}-\frac{y^\alpha}{y^2}\,,
\qquad
Y_{\alpha,ij,kl}=Y_\alpha(x_{ij},x_{kl})\,.
\end{equation}
In this case, all three operators are proper conformal primaries (they
are annihilated by $\hat\kappa_\mu$), and hence
these correlation functions are of the standard form as dictated by
conformal symmetry.
Here, the coefficient depending on the charge and momentum is
\begin{equation}
\coeff^J_p=\frac{1}{\sqrt{J+3}}\frac{-e^{-ip/2}}{\brk{1+e^{i(J+2)p/2}}^{2}}\,.
\label{eq:opecoeff}
\end{equation}
The two-magnon states~\eqref{eq:2magops} are all mutually orthogonal
as well as orthogonal to the vacuum states and all their descendants.
Hence, the coefficients of the two-magnon states in the OPE of two vacuum
operators can simply be extracted from the three- and two-point functions,
\begin{equation}
\bigvev{\op{O}_1(x_1)\,\op{O}_2(x_2)\,\op{O}_3(x_3)}
\xrightarrow{x_2\rightarrow x_3}
\opec^1_{23}(x_{23})\bigvev{\op{O}_1(x_1)\,\op{O}_1(x_2)}
+\dots\,,
\end{equation}
where the ellipsis stands for higher-order terms in $x_{23}$ that
stem from descendants of $\op{O}_1$. The resulting OPE coefficients
for two vacuum operators read
\begin{align}
\opec_{k;k}^{J,p}
=\opec_{k;k}^{J,p,\partial\partial}
&=\coeff^J_pk^2x_{12}^2
\,,\nn\\
\opec_{k;k}^{J,p,VV,(\mu\nu)}
=\opec_{k;k}^{J,p,\partial\partial,(\mu\nu)}
&=\coeff^J_p2k^2x_{12}^{(\mu}x_{12}^{\nu)}
\,,\nn\\
\opec_{k;k}^{J,p,V\partial,1,\mu\nu}
=\opec_{k;k}^{J,p,V\partial,2,\mu\nu}
&=\coeff^J_p2ik^2x_{12}^\mu x_{12}^\nu
\,.
\label{eq:vacOPEcoeff}
\end{align}
%

\paragraph{One Excitation.}

Zero-momentum one-excitation operators transform non-trivially
under $\hat\kappa_\mu$, hence they are not proper conformal
primaries, and their correlation functions will not be
of the familiar form as dictated by conformal symmetry.
By direct computation, one finds the following three-point functions
of two-magnon states (with non-zero
momentum), a one-excitation BPS state $\op{Q}^k_\rho$ and a vacuum
state $\op{Q}^k$:
\begin{align}
\bigvev{\op{O}^{J}_{p,1}\,\op{Q}^{k}_{\rho,2}\,\op{Q}^{k}_3}
&=
\coeff^J_p\sqrt{2}k\frac{x_{23}^{2}}{x_{12}^2x_{13}^2}\bigbrk{\half\bfm{Y}_{\rho,21,23}-Y_{\rho,21,23}}\,,
\nn\\
\bigvev{\op{O}^{J,VV}_{p,(\mu\nu),1}\,\op{Q}^{k}_{\rho,2}\,\op{Q}^{k}_3}
&=
\coeff^J_p\sqrt{2}k\biggbrk{\bfm{Y}_{\rho,21,23}Y_{(\mu,12,13}Y_{\nu),12,13}-\frac{2}{x_{12}^2}I_{12,\rho(\mu}Y_{\nu),12,13}}\,,
\nn\displaybreak[0]\\
\bigvev{\op{O}^{J,\partial\partial}_{p,1}\,\op{Q}^{k}_{\rho,2}\,\op{Q}^{k}_3}
&=
\coeff^J_p\sqrt{2}k\frac{x_{23}^{2}}{x_{12}^2x_{13}^2}\half\bfm{Y}_{\rho,21,23}\,,
\nn\displaybreak[0]\\
\bigvev{\op{O}^{J,\partial\partial}_{p,(\mu\nu),1}\,\op{Q}^{k}_{\rho,2}\,\op{Q}^{k}_3}
&=
\coeff^J_p\sqrt{2}k\,\bfm{Y}_{\rho,21,23}Y_{(\mu,12,13}Y_{\nu),12,13}\,,
\nn\displaybreak[0]\\
\bigvev{\op{O}^{J,V\partial,1}_{p,\mu\nu,1}\,\op{Q}^{k}_{\rho,2}\,\op{Q}^{k}_3}
&=
\coeff^J_p\sqrt{2}ik\biggbrk{\bfm{Y}_{\rho,21,23}Y_{\mu,12,13}Y_{\nu,12,13}-\frac{1}{x_{12}^2}I_{12,\rho\mu}Y_{\nu,12,13}}\,,
\nn\\
\bigvev{\op{O}^{J,V\partial,2}_{p,\mu\nu,1}\,\op{Q}^{k}_{\rho,2}\,\op{Q}^{k}_3}
&=
\coeff^J_p\sqrt{2i}k\biggbrk{\bfm{Y}_{\rho,21,23}Y_{\mu,12,13}Y_{\nu,12,13}-\frac{1}{x_{12}^2}I_{12,\rho\mu}Y_{\nu,12,13}}\,.
\end{align}
Here,
\begin{equation}
\bfm{Y}_{\rho,ij,ik}=\Bigbrk{L\,Y_{\rho,ij,ik}+2k\frac{x_{ik}^\rho}{x_{ik}^2}}\,.
\end{equation}
The resulting OPE coefficients are
\begin{align}
\opec_{k,\rho;k}^{J,p}
&=\coeff^J_p\sqrt{2}k(2k-L+2)\half x_{12,\rho}
\,,\nn\\
\opec_{k,\rho;k}^{J,p,VV,(\mu\nu)}
&=\coeff^J_p\sqrt{2}k
	\Bigbrk{\frac{(2k-L)x_{12}^{(\mu}x_{12}^{\nu)}x_{12,\rho}}{x_{12}^2}+2x_{12}^{(\mu}\eta^{\nu)}_\rho}
\,,\nn\displaybreak[0]\\
\opec_{k,\rho;k}^{J,p,\partial\partial}
&=\coeff^J_p\sqrt{2}k(2k-L)\half x_{12,\rho}
\,,\nn\displaybreak[0]\\
\opec_{k,\rho;k}^{J,p,\partial\partial,(\mu\nu)}
&=\coeff^J_p\sqrt{2}k(2k-L)\frac{x_{12}^{(\mu}x_{12}^{\nu)}x_{12,\rho}}{x_{12}^2}
\,,\nn\displaybreak[0]\\
\opec_{k,\rho;k}^{J,p,V\partial,1,\mu\nu}
&=\coeff^J_p\sqrt{2}ikx_{12}^\nu\Bigbrk{\frac{(2k-L)x_{12}^\mu x_{12,\rho}}{x_{12}^2}+\eta^\mu_\rho}
\,,\nn\\
\opec_{k,\rho;k}^{J,p,V\partial,2,\mu\nu}
&=\coeff^J_p\sqrt{2}ikx_{12}^\nu\Bigbrk{\frac{(2k-L)x_{12}^\mu x_{12,\rho}}{x_{12}^2}+\eta^\mu_\rho}
\,.
\label{eq:1exOPEcoeff}
\end{align}
%

\paragraph{Two Excitations.}

The three-point functions of two-magnon states (with non-zero
momentum) and two one-excitation BPS states read:
\begin{align}
\bigvev{\op{O}^{J}_{p,1}\,\op{Q}^{k}_{\rho,2}\,\op{Q}^{k}_{\sigma,3}}
&=
\coeff^J_p\frac{x_{23}^2}{x_{12}^2x_{13}^2}\biggsbrk{
\biggbrk{\half\bfm{Y}_{\rho,21,23}\bfm{Y}_{\sigma,31,32}-\frac{2k-L}{2x_{23}^2}I_{23,\rho\sigma}}
\nn\\&\mspace{150mu}
-\bfm{Y}_{\rho,21,23}Y_{\sigma,31,32}
-Y_{\rho,21,23}\bfm{Y}_{\sigma,31,32}
-\frac{I_{12,\rho\mu}I_{13}{}^\mu{}_{\sigma}}{x_{23}^2}
}\,,
\nn\displaybreak[0]\\
\bigvev{\op{O}^{J,VV}_{p,(\mu\nu),1}\,\op{Q}^{k}_{\rho,2}\,\op{Q}^{k}_{\sigma,3}}
&=
\coeff^J_p\biggsbrk{
\biggbrk{\bfm{Y}_{\rho,21,23}\bfm{Y}_{\sigma,31,32}-\frac{2k-L}{x_{23}^2}I_{23,\rho\sigma}}
Y_{(\mu,12,13}Y_{\nu),12,13}
\nn\\&\mspace{20mu}
-\bfm{Y}_{\rho,21,23}\frac{2}{x_{13}^2}I_{13,\sigma(\mu}Y_{\nu),13,12}
-\frac{2}{x_{12}^2}I_{12,\rho(\mu}Y_{\nu),12,13}\bfm{Y}_{\sigma,31,32}
-\frac{2I_{12,\rho(\mu}I_{13,\nu)\sigma}}{x_{12}^2x_{13}^2}
}\,,
\nn\displaybreak[0]\\
\bigvev{\op{O}^{J,\partial\partial}_{p,1}\,\op{Q}^{k}_{\rho,2}\,\op{Q}^{k}_{\sigma,3}}
&=
\coeff^J_p\frac{x_{23}^{2}}{x_{12}^2x_{13}^2}
\biggbrk{
\half\bfm{Y}_{\rho,21,23}\bfm{Y}_{\sigma,31,32}
-\frac{2k-L}{2x_{23}^2}I_{23,\rho\sigma}
}
\,,
\nn\displaybreak[0]\\
\bigvev{\op{O}^{J,\partial\partial}_{p,(\mu\nu),1}\,\op{Q}^{k}_{\rho,2}\,\op{Q}^{k}_{\sigma,3}}
&=
\coeff^J_p
\biggbrk{\bfm{Y}_{\rho,21,23}\bfm{Y}_{\sigma,31,32}-\frac{2k-L}{x_{23}^2}I_{23,\rho\sigma}}
Y_{(\mu,12,13}Y_{\nu),12,13}
\,,
\nn\displaybreak[0]\\
\bigvev{\op{O}^{J,V\partial,1}_{p,\mu\nu,1}\,\op{Q}^{k}_{\rho,2}\,\op{Q}^{k}_{\sigma,3}}
&=
\coeff^J_pi\biggsbrk{
\biggbrk{\bfm{Y}_{\rho,21,23}\bfm{Y}_{\sigma,31,32}-\frac{2k-L}{x_{23}^2}I_{23,\rho\sigma}}
Y_{\mu,12,13}Y_{\nu,12,13}
\nn\\&\mspace{100mu}
-\bfm{Y}_{\rho,21,23}\frac{1}{x_{13}^2}I_{13,\sigma\mu}Y_{\nu,13,12}
-\frac{1}{x_{12}^2}I_{12,\rho\mu}Y_{\nu,12,13}\bfm{Y}_{\sigma,31,32}
}\,,
\nn\displaybreak[0]\\
\bigvev{\op{O}^{J,V\partial,2}_{p,\mu\nu,1}\,\op{Q}^{k}_{\rho,2}\,\op{Q}^{k}_{\sigma,3}}
&=
\coeff^J_pi\biggsbrk{
\biggbrk{\bfm{Y}_{\rho,21,23}\bfm{Y}_{\sigma,31,32}-\frac{2k-L}{x_{23}^2}I_{23,\rho\sigma}}
Y_{\mu,12,13}Y_{\nu,12,13}
\nn\\&\mspace{100mu}
-\bfm{Y}_{\rho,21,23}\frac{1}{x_{13}^2}I_{13,\sigma\mu}Y_{\nu,13,12}
-\frac{1}{x_{12}^2}I_{12,\rho\mu}Y_{\nu,12,13}\bfm{Y}_{\sigma,31,32}
}\,.
\end{align}
The resulting OPE coefficients are
\begin{align}
\opec_{k,\rho;k,\sigma}^{J,p}
&=-\coeff^J_p\frac{(2k-L+2)}{2}\Bigbrk{\bfm{I}_{\rho\sigma,12}+\frac{2x_{12,\rho}x_{12,\sigma}}{x_{12}^2}}
\,,\nn\\
\opec_{k,\rho;k,\sigma}^{J,p,VV,(\mu\nu)}
&=-\coeff^J_p
\biggbrk{(2k-L)\frac{x_{12}^{(\mu}}{x_{12}^2}
\biggbrk{2\bigbrk{x_{12,\sigma}\eta^{\nu)}_\rho+x_{12,\rho}\eta^{\nu)}_\sigma}
+x_{12}^{\nu)}\bfm{I}_{\rho\sigma,12}}
+2\eta^{(\mu}_\rho\eta^{\nu)}_\sigma}
\,,\nn\displaybreak[0]\\
\opec_{k,\rho;k,\sigma}^{J,p,\partial\partial}
&=-\coeff^J_p\frac{(2k-L)}{2}\bfm{I}_{\rho\sigma,12}
\,,\nn\displaybreak[0]\\
\opec_{k,\rho;k,\sigma}^{J,p,\partial\partial,(\mu\nu)}
&=-\coeff^J_p(2k-L)
\frac{x_{12}^{(\mu} x_{12}^{\nu)}}{x_{12}^2}\bfm{I}_{\rho\sigma,12}
\,,\nn\displaybreak[0]\\
\opec_{k,\rho;k,\sigma}^{J,p,V\partial,1,\mu\nu}
&=-\coeff^J_pi(2k-L)\frac{x_{12}^\nu}{x_{12}^2}\Bigbrk{\bigbrk{x_{12,\sigma}\eta^\mu_\rho+x_{12,\rho}\eta^\mu_\sigma}+x_{12}^\mu\bfm{I}_{\rho\sigma,12}}
\,,\nn\\
\opec_{k,\rho;k,\sigma}^{J,p,V\partial,2,\mu\nu}
&=-\coeff^J_pi(2k-L)\frac{x_{12}^\nu}{x_{12}^2}\Bigbrk{\bigbrk{x_{12,\sigma}\eta^\mu_\rho+x_{12,\rho}\eta^\mu_\sigma}+x_{12}^\mu\bfm{I}_{\rho\sigma,12}}
\,,
\label{eq:2exOPEcoeff}
\end{align}
where
\begin{equation}
\bfm{I}^{\rho\sigma}(x)
\equiv
I^{\rho\sigma}(x)
+\frac{(2k-L)x^\rho x^\sigma}{x^2}
\,.
\end{equation}
Even though this will not be needed in the subsequent analysis, we
note that for two zero-momentum excitations on one of the two external half-BPS
states, the OPE coefficient of the two-magnon operator $\op{O}^J_p$ reads
\begin{equation}
\opec_{k,(\rho\sigma);k}^{J,p}
=
\coeff^J_pk(2k-L+2)(2k-L)\frac{x_{12,(\rho}x_{12,\sigma)}}{x_{12}^2}
\,.
\end{equation}
%

\paragraph{The Coefficient Function.}

\begin{figure}
\centering
\includegraphics[width=0.5\textwidth]{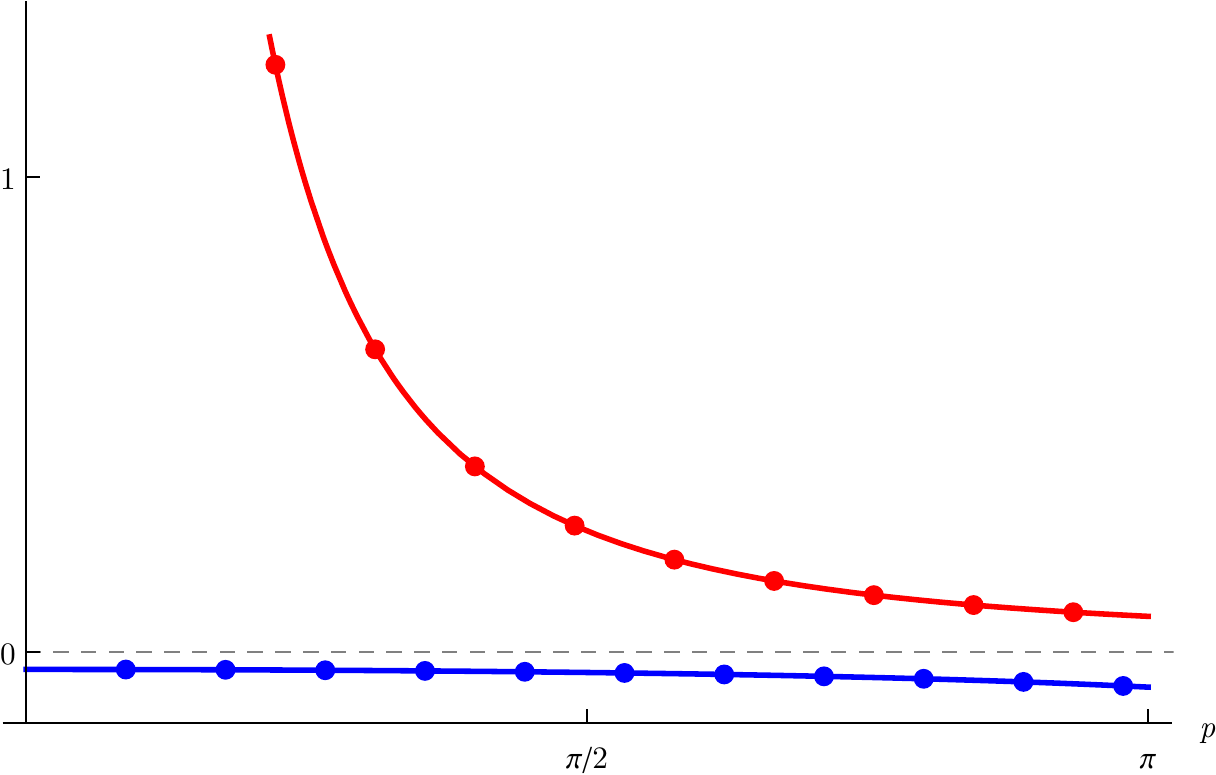}
\caption{The two branches of the OPE coefficient function
$\coeff^J_p$, here displayed for $J=42$. On the lower branch, $p$ is a
multiple of $4\pi/(J+3)$, and therefore $\coeff^J_p\sim
\brk{J+3}^{-1/2}$. On the upper branch, $\coeff^J_p$ becomes large for
small $p$, due to its small denominator. For the smallest two values
$p=2\pi\brc{1,3}/(J+3)$ on the upper branch, the coefficient is off
the chart at $\coeff^J_p\approx\brc{30.6,\,3.41}$.}
\label{fig:coeff1}
\end{figure}
\begin{figure}
\centering
\includegraphicsbox[width=0.53\textwidth]{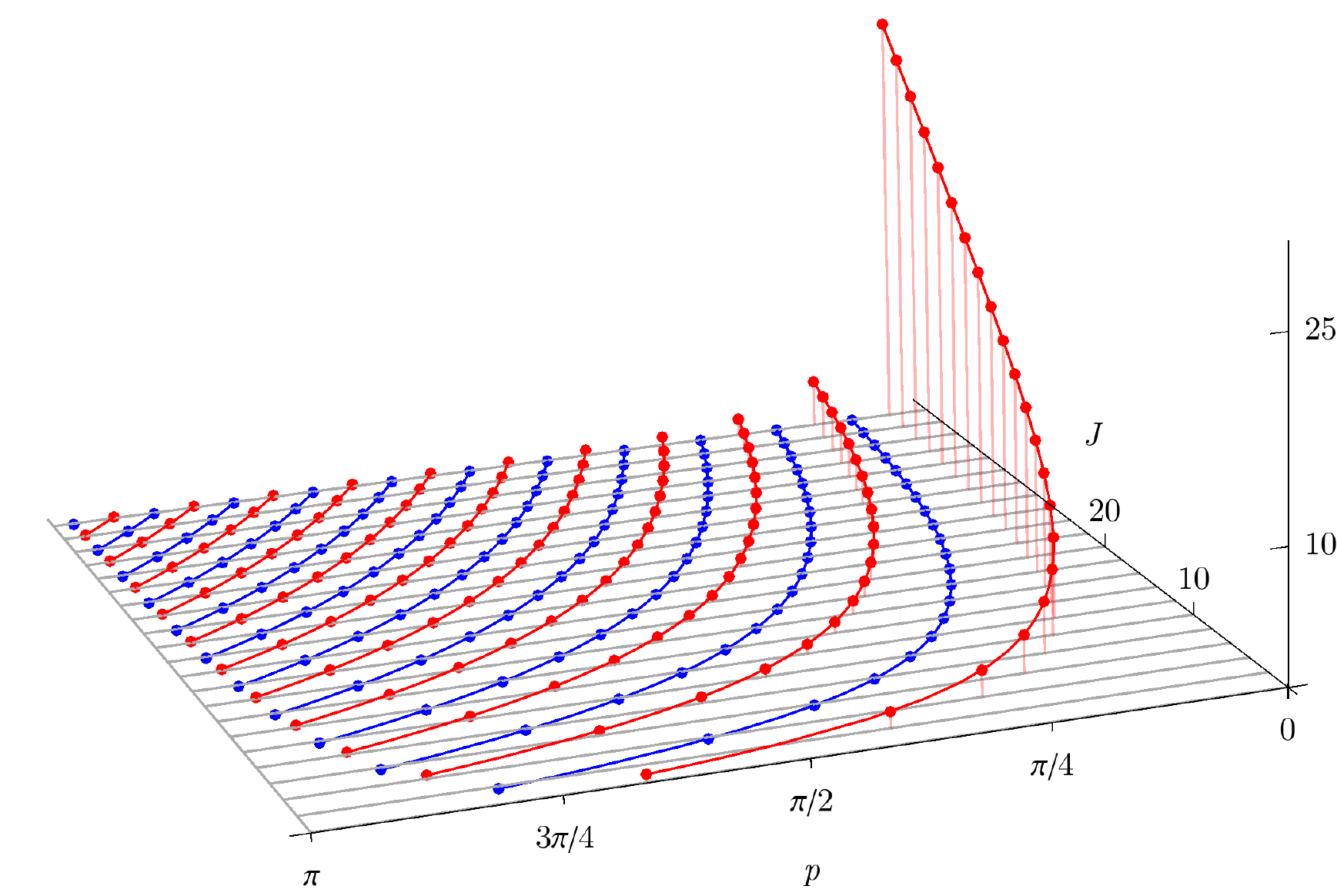}
\hfill
\includegraphicsbox[width=0.43\textwidth]{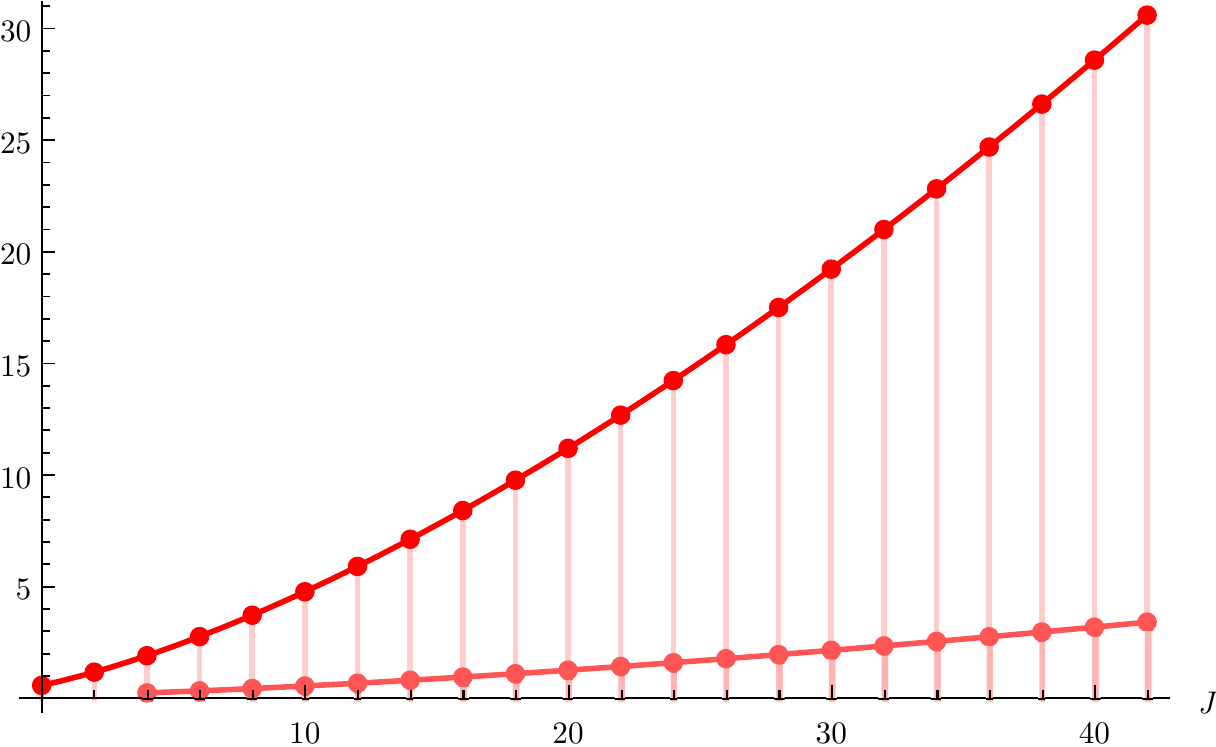}
\caption{The coefficient function $\coeff^J_p$. The plot on the left
runs over all admissible momenta $p$ for $J\leq42$. Every dot stands
for one two-magnon supermultiplet. The ``$-$'' branch
is plotted in red, the ``$+$'' branch is plotted in blue. For each
$J$, the smallest momentum $p=2\pi/(J+3)$ dominates. The plot on the
right shows the two largest values of the coefficient $\coeff^J_p$ on
the dominant ``$-$'' branch ($p=2\pi/(J+3)$ and $p=6\pi/(J+3)$), for
$J\leq42$.}
\label{fig:coeff2}
\end{figure}

Using the on-shell condition~\eqref{eq:momcond}, the OPE coefficent
function~\eqref{eq:opecoeff} can be rewritten as
\begin{equation}
\coeff^J_p=\frac{-1}{2\sqrt{J+3}\bigbrk{\cos(p/2)+e^{ip(J+3)/2}}}\,.
\end{equation}
It has a few interesting properties. First of all, because $p$ is a
multiple of $2\pi/(J+3)$, the term $e^{ip(J+3)/2}$ alternates between
$-1$ and $+1$ when $p$ is varied. The value $+1$ is assumed when $p$
is a multiple of $4\pi/(J+3)$. For fixed $J$, the coefficient
$\coeff^J_p$ therefore has two branches (see~\figref{fig:coeff1}).
Since $p$ takes values between $0$ and $\pi$, the coefficient is small
on the ``$+$'' branch, $\coeff^J_{p,+}\approx-1/4\sqrt{J+3}$. On the
``$-$'' branch, $\coeff^J_p$ becomes large when $p$ is small,
$\coeff^J_{p,-}\approx 4/\sqrt{J+3}\,p^2$.
In fact, for large values of $J$, the smallest momentum
$p=2\pi/(J+3)$ is strongly dominating (see~\figref{fig:coeff2}).
The reason for the alternating behavior of $\coeff^J_p$ is that the
coefficient of the ``local'' two-excitation operator
\begin{equation}
\trb{V_\mu,\phi,\dots,\phi,V_\nu,\phi,\dots,\phi}
\end{equation}
in the OPE of two vacuum (or zero-momentum) operators has a part that
is independent of the distance between the two excitations, and a part that
is proportional to that distance. In the symmetric traceless
two-magnon operator~\eqref{eq:2magops}, the coefficients of these local
operators (i) sum to zero, and (ii) depending on the momentum $p$,
alternate between (almost) odd and even functions of the distance between the two
excitations. Hence, the distance-independent part of the OPE
coefficient drops out, while the linear part is alternatingly
enhanced/suppressed. The same applies to the other types of two-magnon states.
%

\section{Four-Point Correlators}
\label{sec:four-point-corr}

The one-loop correlator of four chiral primary operators
$\op{P}_i=\trb{\brk{X_i\cdot\Phi}^{k}}(x_i)$ of weight $k$, for general
orientations $X_{i}$ with $X_i^2=0$, is given
by~\cite{Arutyunov:2003ae,Drukker:2008pi}
%
\begin{equation}
\vev{\op{P}_1\op{P}_2\op{P}_3\op{P}_4}^{\text{one-loop}}
=
\frac{\lambda}{8\pi^2}
\,I^{(1)}(s,t)
\,k^4
\,\mathcal{R}
\mspace{-20mu}\sum_{\substack{j,\ell,m\geq 0\\j+\ell+m=k-2}}\mspace{-20mu}
\cX^j\cY^\ell\cZ^m
\,,
\label{eq:1loopcorr}
\end{equation}
where
\begin{equation}
\cX=\frac{X_1{\cdot} X_2\,X_3{\cdot} X_4}{x_{12}^2\,x_{34}^2}
\,,
\quad
\cY=\frac{X_1{\cdot} X_3\,X_2{\cdot} X_4}{x_{13}^2\,x_{24}^2}
\,,
\quad
\cZ=\frac{X_1{\cdot} X_4\,X_2{\cdot} X_3}{x_{14}^2\,x_{23}^2}
\end{equation}
are the three possible combinations of propagator structures for
four-points,
\begin{equation}
\cR=s\cX^2+\cY^2+t\cZ^2+\brk{s-t-1}\cY\cZ+\brk{1-s-t}\cX\cZ+\brk{t-s-1}\cX\cY
\end{equation}
is a universal prefactor due to supersymmetry,
\begin{equation}
s=\frac{x_{12}^2\,x_{34}^2}{x_{13}^2\,x_{24}^2}\,,
\qquad
t=\frac{x_{14}^2\,x_{23}^2}{x_{13}^2\,x_{24}^2}
\end{equation}
are the conformally invariant cross-ratios,
and
\begin{align}
I^{(1)}(s,t)
&=
\frac{x_{13}^2\,x_{24}^2}{2\pi^2}
\int\frac{\dd^4x_5}{x_{15}^2\,x_{25}^2\,x_{35}^2\,x_{45}^2}
\end{align}
is the conformal one-loop scalar box integral,
which has been computed in~\cite{Usyukina:1992jd,Usyukina:1993ch}
and is most conveniently written as~\cite{BROWN2004527}
\begin{equation}
I^{(1)}(s,t)
=
-\frac{1}{z-\bar{z}}\Bigbrk{
2\bigbrk{\Li_2(z)-\Li_2(\bar{z})}
-\bigbrk{\Li_1(z)-\Li_1(\bar{z})}\log{s}
}
\,,
\label{eq:phi1}
\end{equation}
where the cross ratios are parametrized in the standard way
\begin{equation}
s=z\bar{z}
\,,\qquad
t=(1+z)(1+\bar{z})
\,.
\end{equation}
We consider Euclidean signature, which means that $z$ and $\bar{z}$
are mutually complex conjugate.

\subsection{Twisted Correlators}

The vacuum operators $\op{Q}$ are deliberately oriented such that
$X_i\cdot X_j=X^\phi\cdot X^\phi=x_{ij}^2$, which sets $\cX=\cY=\cZ=1$, and therefore
$\cR=0$, in accordance with the fact that correlation functions of
vacuum operators are protected:
\begin{equation}
\vev{\op{Q}_1\op{Q}_2\op{Q}_3\op{Q}_4}^{\text{one-loop}}=0
\,.
\end{equation}
Adding zero-momentum excitations $V_\mu$ on top of the
vacuum $\phi$ can be achieved by setting $X_i=X^\phi+\eps_i^\mu
X^{V_\mu}$. To leading order in $\eps_i$, the condition
$X_i\cdot X_i=0$ is still satisfied, and
the four-point correlator expands to
\begin{multline}
\vev{\op{P}_1\op{P}_2\op{P}_3\op{P}_4}
=
\vev{\op{Q}_1\op{Q}_2\op{Q}_3\op{Q}_4}
+\eps_1^{\mu}k\vev{\op{Q}_{\mu,1}\op{Q}_2\op{Q}_3\op{Q}_4}
+\dots
+\eps_1^{\mu}\eps_2^{\nu}k^2\vev{\op{Q}_{\mu,1}\op{Q}_{\nu,2}\op{Q}_3\op{Q}_4}
\\
+\dots
+\eps_1^{\mu}\eps_2^{\nu}\eps_3^{\rho}k^3\vev{\op{Q}_{\mu,1}\op{Q}_{\nu,2}\op{Q}_{\rho,3}\op{Q}_4}
+\dots
+\eps_1^\mu\eps_2^\nu\eps_3^\rho\eps_4^\sigma
	k^4\vev{\op{Q}_{\mu,1}\op{Q}_{\nu,2}\op{Q}_{\rho,3}\op{Q}_{\sigma,4}}
+\dots
\end{multline}
The dependence on $\eps_i^\mu$ enters the
correlator~\eqref{eq:1loopcorr} only through the propagator
structures $\cX$, $\cY$, and $\cZ$, which expand to
\begin{align}
\cX&=
\lrbrk{1+\frac{\sqrt{2}\brk{\eps_1-\eps_2}\cdot x_{12}-\eps_1\cdot\eps_2}{x_{12}^2}}
\lrbrk{1+\frac{\sqrt{2}\brk{\eps_3-\eps_4}\cdot x_{34}-\eps_3\cdot\eps_4}{x_{34}^2}}
\,,\nn\\
\cY&=
\lrbrk{1+\frac{\sqrt{2}\brk{\eps_1-\eps_3}\cdot x_{13}-\eps_1\cdot\eps_3}{x_{13}^2}}
\lrbrk{1+\frac{\sqrt{2}\brk{\eps_2-\eps_4}\cdot x_{24}-\eps_2\cdot\eps_4}{x_{24}^2}}
\,,\nn\\
\cZ&=
\lrbrk{1+\frac{\sqrt{2}\brk{\eps_2-\eps_3}\cdot x_{23}-\eps_2\cdot\eps_3}{x_{23}^2}}
\lrbrk{1+\frac{\sqrt{2}\brk{\eps_1-\eps_4}\cdot x_{14}-\eps_1\cdot\eps_4}{x_{14}^2}}
\,.
\label{eq:XYZexp}
\end{align}
We will denote the expansion coefficients of the universal prefactor
$\cR$ by $\cR^{i,j,\dots}$:
\begin{equation}
\cR
=\sum_{i,j=1}^4\eps_i\eps_j\cR^{i,j}
+\sum_{i,j,k=1}^4\eps_i\eps_j\eps_k\cR^{i,j,k}
+\eps_1\eps_2\eps_3\eps_4\cR^{1,2,3,4}
+\dots
\,.
\end{equation}
The expansion only starts at quadratic
order in the parameters $\eps_i$, hence also correlators
with a single excitation are protected:
\begin{equation}
\vev{\op{Q}_{\mu,1}\op{Q}_2\op{Q}_3\op{Q}_4}^{\text{one-loop}}=0
\,.
\label{eq:4pt1ex}
\end{equation}
At quadratic order, all dependence on the parameters $\eps_i$
is absorbed by the prefactor
$\cR$. Therefore, the sum over $\cX^j\cY^\ell\cZ^m$ trivially evaluates to
$k(k-1)/2$, and one finds for the correlators with two excitations:
\begin{align}
\vev{\op{Q}_{\mu,1}\op{Q}_{\nu,2}\op{Q}_3\op{Q}_4}^{\text{one-loop}}
&=
\frac{\lambda}{8\pi^2}
\,I^{(1)}(s,t)
\,\frac{k^3(k-1)}{2}
\,\cR^{1,2}_{\mu\nu}
\,,
\label{eq:4pt12}
\\
\vev{\op{Q}_{\mu,1}\op{Q}_2\op{Q}_3\op{Q}_{\sigma,4}}^{\text{one-loop}}
&=
\frac{\lambda}{8\pi^2}
\,I^{(1)}(s,t)
\,\frac{k^3(k-1)}{2}
\,\cR^{1,4}_{\mu\sigma}
\,,
\qquad\text{etc.}\,,
\label{eq:4pt14}
\end{align}
where
\begin{multline}
\cR^{1,2,\mu\nu}
=
4\frac{x_{13}^\mu x_{24}^\nu}{x_{13}^2x_{24}^2}
-4s\frac{x_{12}^\mu x_{12}^\nu}{x_{12}^4}
+4t\frac{x_{14}^\mu x_{23}^\nu}{x_{14}^2x_{23}^2}
+2(-1+s+t)\lrbrk{
 \frac{x_{12}^\nu x_{14}^\mu}{x_{12}^2x_{14}^2}
-\frac{x_{12}^\mu x_{23}^\nu}{x_{12}^2x_{23}^2}
}
\\
+2(1+s-t)\lrbrk{
 \frac{x_{12}^\nu x_{13}^\mu}{x_{12}^2x_{13}^2}
-\frac{x_{12}^\mu x_{24}^\nu}{x_{12}^2x_{24}^2}
}
+2(-1+s-t)
\lrbrk{
 \frac{x_{13}^\mu x_{23}^\nu}{x_{13}^2x_{23}^2}
+\frac{x_{14}^\mu x_{24}^\nu}{x_{14}^2x_{24}^2}
}
\,,
\label{eq:R12}
\end{multline}
and all other $\cR^{i,j}$ are given by similar expressions
(see~\appref{sec:details-tensors}).

The correlators
$\vev{\op{Q}_{\mu,1}\op{Q}_{\nu,2}\op{Q}_3\op{Q}_{\sigma,4}}$ and
$\vev{\op{Q}_{\mu,1}\op{Q}_{\nu,2}\op{Q}_{\rho,3}\op{Q}_{\sigma,4}}$
with three or four
excitations are a bit more complicated to obtain
than the previous examples. Upon inserting~\eqref{eq:XYZexp} into the general
formula~\eqref{eq:1loopcorr}, we need to extract the coefficients of
cubic and quartic monomials in the parameters $\eps_i$.
Since the universal prefactor $\cR$ starts at quadratic order, we also
need to expand the sum over propagator structures to quadratic order.
Below, we will consider the double-coincidence limit $\abs{x_{12}},\abs{x_{34}}\ll\abs{x_{23}}$.
In that limit, the expansion of the sum of propagator structures reads:
\begin{multline}
\sum_{\substack{j,\ell,m\geq 0\\j+\ell+m=k-2}}\mspace{-20mu}\cX^j\cY^\ell\cZ^m
=
\frac{k(k-1)}{2}
+\frac{k(k-1)(k-2)}{6}
\Biggsbrk{
\sqrt{2}\lrbrk{\frac{\brk{\eps_1-\eps_2}\cdot x_{12}}{x_{12}^2}+\frac{\brk{\eps_3-\eps_4}\cdot x_{34}}{x_{34}^2}}
\\
+\frac{k-1}{2}\lrbrk{
 \frac{\brk{\eps_1-\eps_2}\cdot x_{12}}{x_{12}^2}
+\frac{\brk{\eps_3-\eps_4}\cdot x_{34}}{x_{34}^2}
}^2
-\frac{\eps_1\cdot I(x_{12})\cdot\eps_2}{x_{12}^2}
-\frac{\eps_3\cdot I(x_{34})\cdot\eps_4}{x_{34}^2}
}
\\
+\order{\eps_i^2}
+\order{\eps_i\eps_j\eps_k}
\,.
\end{multline}
Combining the relevant terms with the coefficients of the prefactor
$\cR$, we find for the one-loop correlator with three excitations, in
the double-coincidence limit:
\begin{multline}
\vev{\op{Q}_{\mu,1}\op{Q}_{\nu,2}\op{Q}_3\op{Q}_{\sigma,4}}^{\text{one-loop}}
=
\frac{\lambda}{8\pi^2}
\,I^{(1)}(s,t)
\,\frac{k^2(k-1)}{2}
\cdot\\\cdot
\lrbrk{
\cR^{1,2,4}_{\mu\nu\sigma}
+\frac{\sqrt{2}(k-2)}{3}
\lrbrk{\frac{x_{12,\mu}}{x_{12}^2}\cR^{2,4}_{\nu\sigma}-\frac{x_{12,\nu}}{x_{12}^2}\cR^{1,4}_{\mu\sigma}+\frac{x_{34,\sigma}}{x_{34}^2}\cR^{1,2}_{\mu\nu}}
}\,,
\label{eq:4pt3ex}
\end{multline}
and for the one-loop correlator with four excitations:
\begin{multline}
\vev{\op{Q}_{\mu,1}\op{Q}_{\nu,2}\op{Q}_{\rho,3}\op{Q}_{\sigma,4}}^{\text{one-loop}}
=
\frac{\lambda}{8\pi^2}
\,I^{(1)}(s,t)
\,\frac{k(k-1)}{2}
\cdot\\\cdot
\Biggbrk{
\cR^{1,2,3,4}_{\mu\nu\rho\sigma}
+\frac{(k-2)}{3}
\Biggsbrk{
\sqrt{2}\lrbrk{
 \frac{x_{12,\mu}}{x_{12}^2}\cR^{2,3,4}_{\nu\rho\sigma}
-\frac{x_{12,\nu}}{x_{12}^2}\cR^{1,3,4}_{\mu\rho\sigma}
+\frac{x_{34,\sigma}}{x_{34}^2}\cR^{1,2,3}_{\mu\nu\rho}
}
\mspace{150mu}
\\
+(k-1)\biggbrk{
-\frac{x_{12,\mu}x_{12,\nu}}{x_{12}^4}\cR^{3,4}_{\rho\sigma}
+\frac{x_{12,\mu}x_{34,\rho}}{x_{12}^2x_{34}^2}\cR^{2,4}_{\nu\sigma}
-\frac{x_{12,\mu}x_{34,\sigma}}{x_{12}^2x_{34}^2}\cR^{2,3}_{\nu\rho}
-\frac{x_{12,\nu}x_{34,\rho}}{x_{12}^2x_{34}^2}\cR^{1,4}_{\mu\sigma}
\\
+\frac{x_{12,\nu}x_{34,\sigma}}{x_{12}^2x_{34}^2}\cR^{1,3}_{\mu\rho}
-\frac{x_{34,\rho}x_{34,\sigma}}{x_{12}^4}\cR^{1,2}_{\mu\nu}
}
-\frac{I_{12,\mu\nu}}{x_{12}^2}\cR^{3,4}_{\rho\sigma}
-\frac{I_{34,\rho\sigma}}{x_{34}^2}\cR^{1,2}_{\mu\nu}
}
}\,.
\label{eq:4pt4ex}
\end{multline}
The lengthier expressions for the coefficients $\cR^{i,j,k}$ and $\cR^{1,2,3,4}$
are provided in~\appref{sec:details-tensors}.

\section{OPE Analysis}

We want to compute the leading contribution to the OPE expansion
of various one-loop four-point correlators. That is, we consider the
double coincidence limit (OPE limit) $\abs{x_{12}},\abs{x_{34}}\ll\abs{x_{23}}$,
or equivalently
$s\to0$, $t\to1$. In terms of the variables $z$, $\bar{z}$, the limit
is attained for $\abs{z}\to0$, upon which the one-loop box integral~\eqref{eq:phi1}
expands to
\begin{equation}
I^{(1)}(s,t)
=\log(s)-2+\order{\sqrt{s}}
\,.
\end{equation}
In the following, we will focus on the leading term $\log(s)$ only.
In the OPE decomposition, this term is completely captured by the anomalous
dimensions of the internal operators: For an operator with scaling dimension
\begin{equation}
\Delta=\Delta_0+\lambda\Delta_1+\order{\lambda^2}\,,
\end{equation}
the two-point function expands to
\begin{equation}
\frac{1}{x^{2\Delta}}
=
\frac{1}{x^{2\Delta_0}}\brk{1-2\lambda\Delta_1\log\abs{x}}
+\order{\lambda^2}
\,.
\label{eq:twoptlog}
\end{equation}
Considering the OPE sum of any four-point correlator,
\begin{equation}
\bigvev{\op{O}_1(x_1)\op{O}_2(x_2)\op{O}_3(x_3)\op{O}_4(x_4)}
=
\sum_M\opec_{12}^M(x_{12})\,\vev{\op{O}_M(x_2)\op{O}_M(x_3)}\,\opec_{34}^M(x_{34})
\,,
\label{eq:opesum}
\end{equation}
it is clear that, at one-loop order, every operator $\op{O}_M$ with a non-vanishing
anomalous dimension will contribute a $\log\abs{x_{23}}$ term. Now,
the double coincidence limit is equivalent to letting
$\abs{x_{23}}\to\infty$, which implies the identification
\begin{equation}
\log(s)=-4\log\abs{x_{23}}
\,.
\label{eq:logident}
\end{equation}
Moreover, since $\vev{\op{O}_M(x_2)\op{O}_M(x_3)}\sim\abs{x_{23}}^{-\Delta_0}$, the leading term in
the OPE sum in the $\abs{x_{23}}\to\infty$ limit stems from operators $\op{O}_M$ with lowest classical
dimension $\Delta_0$. Combining~\eqref{eq:twoptlog},
\eqref{eq:opesum}, and~\eqref{eq:logident}, one finds that the leading
$\log(s)$ part of any one-loop four-point correlator is given by
\begin{multline}
\bigvev{\op{O}_1(x_1)\op{O}_2(x_2)\op{O}_3(x_3)\op{O}_4(x_4)}\supup{one-loop}
=
\\
\frac{\lambda}{2}\log(s)
\sum_M\Delta_{M,1}\lrbrk{\opec_{12}^M(x_{12})\,\vev{\op{O}_M(x_2)\op{O}_M(x_3)}\,\opec_{34}^M(x_{34})}\supup{class.}
+\dots\,,
\label{eq:opesum1}
\end{multline}
where the sum runs over operators $\op{O}_M$ of the smallest classical
dimension $\Delta_0$ that have a non-vanishing anomalous dimension
$\Delta_1$, ``class.'' stands for the classical part, and the ellipsis
stands for terms that are power-suppressed in $s$ or $\log(s)$.

Specializing to operators made of the twisted fields $\phi(x)$,
$V_\mu(x)$, and $B(x)$, the operators with lowest twisted scaling
dimension and
non-vanishing anomalous dimension are exactly the two-magnon
states~\eqref{eq:2magops}, and therefore the sum in~\eqref{eq:opesum1}
runs over exactly those states. Below, we will compute the OPE sum for
several cases.

The full two-magnon supermultiplet comprises $13$ types of
states~\cite{Beisert:2002tn}. Four of those have only fermionic
excitations, and since we will only consider bosonic excitations on
the external states, their classical OPE coefficients will always vanish. Hence
we can neglect these types of states. Another three types of states
are antisymmetric under exchange of their two excitations. Since we
will not consider antisymmetric excitations on the external states,
these types of operators can also be neglected. What remains are the
six types of operators with bosonic excitations listed
in~\eqref{eq:2magops}.

In the OPE expansion of four scalar operators of length $k$,
generically all states with lengths $L=\brc{2,\dots,2k}$ contribute.
Of the six types of operators in the two-magnon
supermultiplet~\eqref{eq:2magops}, three types have length $L=J+2$,
while the other three types have length $L=J+4$. At a given length
$L$, six types of operators contribute; they belong to two different
supermultiplets with $J=L-2$ and $J=L-4$. Generically, complete
multiplets contribute to the OPE. The exception is the extremal case
$L=2k$, where only one half of the multiplet with $J=L-2$ contributes.

In the following, we will compute the OPE sum of the right-hand side
of~\eqref{eq:opesum1} for several example correlators, using the OPE
coefficients obtained in~\secref{sec:two-point-three}. An overview of
the results is shown in~\tabref{tab:results}.
\begin{table}
\centering
\begin{tabular}{ccccc}
\toprule
Correlator \&                   & Correlator         & Match with full  & Remainder                & Tensor structure \\
OPE channel                     & expression         & supermultiplets  & from half SM             & from single SM \\
\midrule
\includegraphicsbox{figcorr00}  & 0                  & $\checkmark$     & 0                        & $\checkmark$ \\[1.5ex]
\includegraphicsbox{figcorr10}  & 0                  & $\checkmark$     & \eqref{eq:1exremainder}  & $\checkmark$ \\[1.5ex]
\includegraphicsbox{figcorr20}  & \eqref{eq:4pt12}   & $\checkmark$     & \eqref{eq:20extremal}    & $\checkmark$ \\[1.5ex]
\includegraphicsbox{figcorr11}  & \eqref{eq:4pt14}   & $\checkmark$     & \eqref{eq:11extremal}    & $\checkmark$ \\[1.5ex]
\includegraphicsbox{figcorr21}  & \eqref{eq:4pt3ex}  & $\checkmark$     & \eqref{eq:21extremal}    & $\text{\small\sffamily X}$ \\[1.5ex]
\includegraphicsbox{figcorr22}  & \eqref{eq:4pt4ex}  & $\checkmark$     & \eqref{eq:22extremal}    & $\text{\small\sffamily X}$ \\
\bottomrule
\end{tabular}
\caption{Overview of results of the OPE analysis for the various
four-point correlators. Black dots stand for vacuum operators
$\op{Q}$, white dots stand for excited operators $\op{Q}_\mu$.
Column~3 indicates the matching of the OPE sum over full
two-magnon multiplets (with $0\leq J\leq 2k-4$) with the
OPE limit of the known one-loop correlator (column~2).
Column~4 lists references to the spurious remainders from
extremal multiplets (with $J=2k-2$). The last column indicates whether
the OPE sum over a single supermultiplet already produces the right
tensor structure of the correlator.}
\label{tab:results}
\end{table}
In all cases, we find that the double-coincidence limit of the known
four-point correlators (\secref{sec:four-point-corr}) is correctly
reproduced by the OPE sum over \emph{complete} two-magnon
supermultiplets. The operators of the extremal half-multiplet at
$J=2k-2=L-2$ produce a spurious contribution that has to be canceled
by double-trace states, on which we comment in~\secref{sec:note-double-trace}.
Unless the whole four-point correlator vanishes, the spurious terms
are suppressed in the limit of large weight $k$.

\subsection{Four Vacua}

We start with the correlator of four vacuum states. It is easy to see
from the two-point functions~\eqref{eq:twopts} and OPE
coefficients~\eqref{eq:vacOPEcoeff} that the contributions of
the operators
$\op{O}^J_p$,
$\op{O}^{J,\partial\partial}_{p,(\mu\nu)}$, and
$\op{O}^{J,V\partial,1}_{p,\mu\nu}$ as well as the contributions of
the operators
$\op{O}^{J,\partial\partial}_p$,
$\op{O}^{J,VV}_{p,(\mu\nu)}$, and
$\op{O}^{J,V\partial,2}_{p,\mu\nu}$ separately sum up to zero. Hence the OPE
limit vanishes at one-loop order. This is expected,
since the vacuum correlator $\vev{\op{Q}_1\op{Q}_2\op{Q}_3\op{Q}_4}$ is
protected.

\subsection{One Excitation}

Next, consider the four-point correlator of one
single-excitation operator with three vacuum operators,
$\vev{\op{Q}_{\mu,1}\op{Q}_2\op{Q}_3\op{Q}_4}$. For every $J$ and
$p$, using the two-point functions~\eqref{eq:twopts} and the OPE
coefficients~\eqref{eq:vacOPEcoeff,eq:1exOPEcoeff}, one finds
that the OPE expansion can be split into a part proportional to
$(2k-L)$ and a remainder. All pieces proportional to $(2k-L)$ cancel
among states with $L=J+2$, and similarly among states with $L=J+4$
within a single multiplet. The remaining terms also cancel within a
single multiplet, but in combinations of terms with $L=J+2$ and
$L=J+4$. As a result, each full supermultiplet contributes zero. This
is the expected result, since the full one-excitation correlator is
protected~\eqref{eq:4pt1ex}. But
the extremal case $L=2k$ produces a remainder, from the half-multiplet
with $J=L-2=2k-2$. The contribution of this remainder to the classical
OPE reads
\begin{equation}
\sum_{p}
\brk{\coeff^{2k-2}_p}^2
\,2^{3/2}k^3
x_{12}^\alpha
\bigbrk{-I_{23,\mu\beta}I_{23,\alpha\gamma}+\half\eta_{\mu\alpha}\eta_{\beta\gamma}}x_{23}^{-4}
x_{34}^\beta x_{34}^\gamma
\,.
\end{equation}
Adding the one-loop anomalous dimension~\eqref{eq:anomdim},
the sum over momenta gives
\begin{equation}
\sum_p\brk{\coeff^{2k-2}_p}^2\Delta_1(p)
=\frac{L}{8\pi^2}=\frac{J+2}{8\pi^2}=\frac{k}{4\pi^2}
\,.
\label{eq:momsum}
\end{equation}
Hence, the contributions of the extremal
half-multiplets sum to
\begin{equation}
\frac{\lambda}{2}\log(s)
\frac{k^4}{\sqrt{2}\,\pi^2}
\,x_{12}^\alpha
\bigbrk{-I_{23,\mu\beta}I_{23,\alpha\gamma}+\half\eta_{\mu\alpha}\eta_{\beta\gamma}}x_{23}^{-4}
x_{34}^\beta x_{34}^\gamma
\,.
\label{eq:1exremainder}
\end{equation}
Since the one-excitation correlator vanishes~\eqref{eq:4pt1ex}, this
result needs to be canceled by another contribution. At leading order
in $1/\abs{x_{23}}$, the supermultiplets~\eqref{eq:2magops} exhaust
the single-trace contributions, but it is well-known that extremal
states with $L=2k$ mix with double-trace operators. Adding the
contribution of those operators should cancel the
term~\eqref{eq:1exremainder}. See the discussion in~\secref{sec:note-double-trace}
below.

\subsection{Two Excitations on One Side}

Combining the OPE coefficients~\eqref{eq:1exOPEcoeff}
and~\eqref{eq:2exOPEcoeff} with the
two-point functions~\eqref{eq:twopts}, one finds that the
contributions of individual operators to the OPE of the correlator
$\vev{\op{Q}_{\mu,1}\op{Q}_{\nu,2}\op{Q}_3\op{Q}_4}$ are quadratic
in $(2k-L)$. The sum of contributions within each half-multiplet (with
$L=J+2$ or $L=J+4$) is only linear in $(2k-L)$, the quadratic piece
cancels. In the sum over each full supermultiplet, all dependence on
$(2k-L)$ cancels (except for overall factor $\brk{\coeff^J_p}^2k^2$).
The contribution of all full supermultiplets to the OPE reads
\begin{equation}
\sum_{\substack{J=0\\J\,\text{even}}}^{2k-4}\sum_p
-\brk{\coeff^J_p}^22k^2
\Bigbrk{
I_{12,\mu\gamma}I_{23}{}^\gamma{}_{\alpha}I_{23,\nu\beta}
+\frac{x_{12,\mu}x_{12,\nu}}{x_{12}^2}\eta_{\alpha\beta}
+(\mu\leftrightarrow\nu)
}
x_{23}^{-4}x_{34}^\alpha x_{34}^\beta
\,.
\end{equation}
Adding the anomalous dimension, the sum over multiplets
(momenta $p$ and weights $J$) gives
\begin{equation}
\sum_{\substack{J=0\\J\,\text{even}}}^{2k-4}\sum_p\brk{\coeff^J_p}^2\Delta_1(p)
=\frac{k(k-1)}{8\pi^2}
\,.
\label{eq:Jpsum}
\end{equation}
Hence the contributions of all complete two-magnon supermultiplets
to the one-loop four-point correlator sum to
\begin{equation}
\frac{\lambda}{2}\log(s)
\frac{k^3(k-1)}{8\pi^2}
\Bigsbrk{-2\Bigbrk{
I_{12,\mu\gamma}I_{23}{}^\gamma{}_{\alpha}I_{23,\nu\beta}
+\frac{x_{12,\mu}x_{12,\nu}}{x_{12}^2}\eta_{\alpha\beta}
+(\mu\leftrightarrow\nu)
}
x_{23}^{-4}x_{34}^\alpha x_{34}^\beta
}
\,.
\label{eq:12opesum}
\end{equation}
The term in the square brackets exactly equals the prefactor
$\cR^{1,2}_{\mu\nu}$~\eqref{eq:R12limit} in the double-coincidence limit
$\abs{x_{12}},\abs{x_{34}}\ll\abs{x_{23}}$. Comparing
with~\eqref{eq:4pt12}, we therefore find that~\eqref{eq:12opesum}
reproduces the limit of the known four-point correlator.

However, there is also the extremal half supermultiplet at $L=2k=J+2$,
which gives the following non-trivial contribution to the classical OPE:
\begin{equation}
\sum_p
-\brk{\coeff^{2k-2}_p}^2k^2
\frac{\eta_{\mu\nu}x_{34}^2}{x_{23}^4}\,.
\end{equation}
Adding the anomalous dimension $\Delta_1(p)$ and summing over
momenta~\eqref{eq:momsum} results in the following extra contribution:
\begin{equation}
\frac{\lambda}{2}\log(s)
\lrbrk{
-\frac{k^3}{4\pi^2}
\frac{\eta_{\mu\nu}x_{34}^2}{x_{23}^4}
}\,.
\label{eq:20extremal}
\end{equation}
Again, this should be canceled by contributions from the relevant
double-trace operators.

\subsection{One Excitation on Each Side}

Next, consider the correlator $\vev{\op{Q}_{\mu,1}\op{Q}_2\op{Q}_3\op{Q}_{\sigma,4}}$.
The analysis parallels the previous case.
Again, each individual operator contribution to the OPE is quadratic in
$(2k-L)$, the contribution of each half-multiplet is linear in
$(2k-L)$, and the contribution of each full supermultiplet is independent
of $(2k-L)$. The contributions from full supermultiplets to the
classical OPE give
\begin{equation}
\sum_{\substack{J=0\\J\,\text{even}}}^{2k-4}\sum_p
-\brk{\coeff^J_p}^24k^2
x_{12}^\alpha
\Bigbrk{
\half I_{23,\alpha\beta}\bigbrk{I_{12,\mu\gamma}I_{23}{}^\gamma{}_{\sigma}+I_{23,\mu\gamma}I_{34}{}^\gamma{}_{\sigma}-2I_{23,\mu\sigma}}
+I_{23,\mu\beta}I_{23,\sigma\alpha}
+\eta_{\mu\alpha}\eta_{\sigma\beta}
}
x_{23}^{-4}x_{34}^\beta
\,.
\end{equation}
Inserting the one-loop anomalous dimension and performing the state
sums yields
\begin{multline}
\frac{\lambda}{2}\log(s)
\frac{k^3(k-1)}{8\pi^2}
\Bigsbrk{
-4x_{12}^\alpha
\Bigbrk{
\half I_{23,\alpha\beta}\bigbrk{I_{12,\mu\gamma}I_{23}{}^\gamma{}_{\sigma}+I_{23,\mu\gamma}I_{34}{}^\gamma{}_{\sigma}-2I_{23,\mu\sigma}}
\\
+I_{23,\mu\beta}I_{23,\sigma\alpha}
+\eta_{\mu\alpha}\eta_{\sigma\beta}
}
x_{23}^{-4}x_{34}^\beta
}
\label{eq:14opesum}
\end{multline}
for the one-loop OPE.
The expression in the square brackets equals the leading part of
$\cR^{1,4}_{\mu\sigma}$~\eqref{eq:R14limit} in the double-coincidence limit
$\abs{x_{12}},\abs{x_{34}}\ll\abs{x_{23}}$, and
hence~\eqref{eq:14opesum} equals the leading part of the known
correlator~\eqref{eq:4pt14} in that limit.

Again, there is an extremal half-multiplet at $2k=L=J+2$, which
produces the following contribution to the classical OPE:
\begin{equation}
\sum_p
\brk{\coeff^{2k-2}_p}^22k^2
x_{12}^\alpha
\bigbrk{
I_{23,\mu\sigma}I_{23,\alpha\beta}
-\eta_{\mu\alpha}\eta_{\sigma\beta}
}
x_{23}^{-4}x_{34}^\beta
\,.
\end{equation}
Via~\eqref{eq:momsum}, this gives rise to the spurious contribution
\begin{equation}
\frac{\lambda}{2}\log(s)
\frac{k^3}{2\pi^2}
x_{12}^\alpha
\bigbrk{
I_{23,\mu\sigma}I_{23,\alpha\beta}
-\eta_{\mu\alpha}\eta_{\sigma\beta}
}
x_{23}^{-4}x_{34}^\beta
\label{eq:11extremal}
\end{equation}
to the one-loop OPE. Again, this should be canceled by
the relevant double-trace contributions.

\subsection{Three Excitations}

In the case of the three-excitation correlator
$\vev{\op{Q}_{\mu,1}\op{Q}_{\nu,2}\op{Q}_{3}\op{Q}_{\sigma,4}}$,
summing over all contributions of a single supermultiplet yields a
lengthy expression. In the contributions of full supermultiplets to
the classical OPE
\begin{equation}
\sum_{\substack{J=0\\J\,\text{even}}}^{2k-4}\sum_p
\brk{\coeff^J_p}^2k^2\bigbrk{\dots}
\end{equation}
the term in parantheses can be constructed from the OPE
coefficients~\eqref{eq:1exOPEcoeff,eq:2exOPEcoeff} and two-point
functions~\eqref{eq:twopts}. It is a long expression that depends both on $k$
(quadratically) and $J$ (linearly). Nonetheless, inserting the
one-loop anomalous dimension $\Delta_1$ and performing the sums over
momenta $p$ and charges $J$ exactly yields the expected
expression~\eqref{eq:4pt3ex}.

As in the previous cases, the extremal half supermultiplet contributes
a spurious remainder, which takes the surprisingly simple form
\begin{equation}
\frac{\lambda}{2}\log(s)
\sum_p\brk{\coeff^{2k-2}_p}^2\Delta_1(p)\bigbrk{\dots}
=
\frac{\lambda}{2}\log(s)
\frac{k^2}{2^{3/2}\pi^2}
\frac{\eta_{\mu\nu}x_{34,\sigma}}{x_{23}^4}
\,,
\label{eq:21extremal}
\end{equation}
and wich should be canceled by the relevant double-trace contributions.

\subsection{Four Excitations}

The contributions to the OPE of the four-excitation correlator
$\vev{\op{Q}_{\mu,1}\op{Q}_{\nu,2}\op{Q}_{\rho,3}\op{Q}_{\sigma,4}}$
are yet more complicated functions. In the sum over full supermultiplets
\begin{equation}
\sum_{\substack{J=0\\J\,\text{even}}}^{2k-4}\sum_p
\brk{\coeff^J_p}^2k\bigbrk{\dots}
\end{equation}
the term in parentheses is again a long expression (constructed
from~\eqref{eq:2exOPEcoeff,eq:twopts}) that depends
quadratically on both $k$ and $J$. Yet, performing the sums yields
exactly the wanted result~\eqref{eq:4pt4ex}. Also in this case, the
spurious term from the extremal half supermultiplet at $J=2k-2$ is
remarkably simple:
\begin{equation}
\frac{\lambda}{2}\log(s)
\sum_p\brk{\coeff^{2k-2}_p}^2\Delta_1(p)\bigbrk{\dots}
=
\frac{\lambda}{2}\log(s)
\frac{k}{4\pi^2}
\frac{\eta_{\mu\nu}\eta_{\rho\sigma}}{x_{23}^4}
\,.
\label{eq:22extremal}
\end{equation}
Again, this term should be canceled by contributions from double-trace
operators.

\subsection{Note on Double-Trace Operators}
\label{sec:note-double-trace}

In all examples above, we have seen that the leading term in
the double-coincidence limit of the one-loop correlator is correctly
reproduced by summing over complete two-magnon supermultiplets in the
OPE channel, neglecting the extremal half-supermultiplet with charge
$J=2k-2$. Including the extremal half-multiplet in the OPE sum leads
to a spurious contribution that needs to be canceled by further
operators. Since the two-magnon operators~\eqref{eq:2magops} exhaust
the single-trace spectrum at $\hat\Delta_0=2$ and $\Delta_1\neq0$,
those further states have to be multi-trace states. In fact, it is
well-known that single-trace and double-trace operators mix at
subleading order in
$1/\Nc$~\cite{Constable:2002hw,Beisert:2002bb,Constable:2002vq}. That
is, single-trace operators obtain double-trace corrections that are
suppressed by a factor of $1/\Nc$. In extremal three-point functions,
each trace of the double-trace operator can separately contract with
the two other external operators. Such terms are enhanced by a factor
$\Nc$ that cancels the $1/\Nc$ suppression. Hence they affect
extremal three-point functions at leading order in $1/\Nc$. The
double-trace corrections to our two-magnon states~\eqref{eq:2magops}
carry two excitations as well. There are three cases to distinguish:
\begin{enumerate}
\item[\bfseries (i)]
both excitations sit on the same trace factor, with a non-zero
momentum,
\item[\bfseries (ii)]
both excitations sit on the same trace factor, with
zero momentum,
\item[\bfseries (iii)]
each of the two trace factors carries one of the
two excitations.
\end{enumerate}
In the first case, the excited trace factor is itself
one of the two-magnon states~\eqref{eq:2magops}. In case (ii) and
(iii), each of the excited trace factors is a descendant of the vacuum
operator. Two-magnon states with non-zero momentum contract to zero
with the external operators considered above, hence corrections of
type (i) do not contribute to the OPE. We are thus looking for a
double-trace correction of type (ii) or (iii) to a two-magnon
state~\eqref{eq:2magops}. Such corrections do indeed exist: For
example, the singlet state $\op{O}_p^J$ gets a correction of type
(iii) (see equation (3.27) in~\cite{Beisert:2002bb}).

There are further contributions to the OPE due to double-trace
operators: While double-trace operators of type (ii) and (iii) have vanishing
anomalous dimension at leading order in the planar limit, they acquire single-trace
corrections at subleading order in $1/\Nc$, which can lead to
$1/\Nc^2$ corrections to their anomalous dimensions. Combined with the
$\Nc$ enhancement of extremal three-point functions with double-trace
operators, also these corrections can affect the four-point OPE at leading
order in $1/\Nc$. Indeed, double-trace operators of type (iii) get
corrected by two-magnon states with non-zero momenta (see equation
(3.16) in~\cite{Beisert:2002bb}), which shows that this effect indeed
occurs.

Our results show that the contributions from extremal states at charge
$J=2k-2$ need to be canceled by double-trace contributions of the type
described above. On a technical level, computing all such double-trace contributions
explicitly is not entirely straightforward, because the
orthogonalization procedure for zero-momentum states is non-trivial
already in the untwisted case~\cite{Beisert:2002bb}, and is further
complicated by the twisting. Nonetheless, it would be
desirable to compute the double-trace contributions in order
to confirm their cancellation. Here, we take the standpoint that the extremal contributions
\eqref{eq:1exremainder,eq:20extremal,eq:11extremal,eq:21extremal,eq:22extremal}
provide predictions for the sum of all double-trace contributions
to the various OPEs.

\section{Conclusions}

We have considered four-point correlators of single-trace scalar
operators that are organized in an excitation picture around specific
vacuum operators. Both the vacuum operators and the excitations are
related to the familiar $\alg{psu}(2,2|4)$ spin chain picture via a
twisting procedure that ties the orientation of the vacuum in the
internal $\grp{S}^5$ to the spacetime location. The motivation for
this choice of operators is that any number of such operators is
preserved by two common supercharges $\alg{Q}^\pm$, and that all other
operators arrange themselves in terms of excitations on top of the
vacuum operators that fall in representations of a common $\alg{q}(3)$
symmetry algebra. These properties make the twisted operators promising candidates for an
integrability-based treatment of higher-point correlation functions.

In this study, we have barely scratched the surface of these
correlation functions: We have only considered zero-momentum
excitations, and in the OPE decomposition, we have restricted
ourselves to the leading non-trivial terms. Nevertheless, we can make
a few non-trivial observations:

For every correlator with up to two excitations that we
considered, we find that each complete two-magnon
supermultiplet separately contributes the correct tensor structure to
the one-loop correlator in the double-coincident limit.
In particular, for protected correlators, the contributions from each
two-magnon supermultiplet separately sum up to zero.
This appears to be a non-trivial fact. It could as
well have been that the correct tensor structure is only obtained
after summing over multiplets.

For every correlator that we considered, performing the sum over
complete two-magnon multiplets (momenta $p$ and charges $J\leq 2k-4$) exactly
reproduces the OPE limits of the known four-point correlators. In
particular, this sum over complete supermultiplets does not include
states from the multiplet at ``extremal'' charge $J=2k-2$, even though
these states \emph{do} contribute non-trivially to the OPE. The fact
that those states need to be discarded in order to recover the known
correlators shows that their contributions need to be canceled by
further states. Since the two-magnon states we considered are the only
single-trace states that can contribute at leading order in the
double-coincidence limit, those further states have to be double-trace
states. These are known to undergo a $1/\Nc$ mixing with single-trace
operators in extremal OPE coefficients (three-point functions),
and thus do contribute to the OPE at leading order in
$1/\Nc$. Even though we have not computed the double-trace
contributions, our results show that they have to cancel the
contributions of the extremal single-trace supermultiplet at $J=2k-2$.
Turning the logic around, our results constitute predictions
for the double-trace contributions to the various OPEs.

In general, four- and higher-point functions have an intricate
spacetime dependence, both through complicated tensor structures and
through non-trivial functional dependence on conformally invariant
cross ratios. Perhaps one of the strongest advantages of the twisted
correlators is that all spacetime dependence is directly coupled to
the magnons (excitations) on the external operators. It can be
seen directly from the contraction rules~\eqref{eq:contractions} that inserting
excitations on top of the vacuum operators injects spacetime factors
into the free-field contractions. Correspondingly, the respective
spacetime factors emerge from the general formula~\eqref{eq:1loopcorr}
for half-BPS operators, as in the examples~\eqref{eq:4pt12},
\eqref{eq:4pt14}, \eqref{eq:4pt3ex}, and~\eqref{eq:4pt4ex}.

The OPE of vacuum operators $\trb{\phi\dots\phi}$ alone already
is an interesting object. The set of vacuum operators is not closed
under the OPE, as exemplified by the non-trivial OPE
coefficients~\eqref{eq:vacOPEcoeff} with two-magnon states considered
here. Since all correlation functions of vacuum operators are
protected, and evaluate to constants, this implies non-trivial
cancellations among all operators that flow in the OPE.

\subsection*{Outlook}

We have only taken the very first steps in the study of higher-point
correlation functions of twisted operators, which leaves a lot of room
for further explorations. The possibilities include:

\medskip
\noindent
It would be very interesting to see whether extremal and
double-trace contributions continue to cancel out also for more general
operators (with non-zero-momentum excitations), and beyond leading order
in the double-coincidence limit. If this holds, it would mean that one
could entirely avoid to resolve the mixing with double-trace
operators. To this end, it would be interesting to compute the
double-trace contributions explicitly for some examples.

It would be very desirable to promote the heuristic results presented here to a
more systematic description of four-point (and perhaps higher-point)
functions, based on the excitation picture over twisted vacua. The
only ingredients should be the magnon rapidities and flavors of the
excitations on the four external operators, the two-magnon S-matrix,
as well as general symmetry principles. In particular, also the
spacetime dependence should be captured completely; since it is
directly tied to the external magnons, it can perhaps be absorbed in a
redefined two-body S-matrix.

In order to understand the above points, one should obviously
extend the present analysis beyond half-BPS operators by including
excitations with non-zero momenta on the external states.

Excitations on top of the vacuum operators $\trb{\phi\dots\phi}$
organize in representations of a common $\alg{q}(3)$ symmetry algebra.
It would be very interesting to study the
consequences of this symmetry.

We have used the lowest-lying non-protected states to
reconstruct the leading logarithmic term of the one-loop correlators
in the double-coincidence limit. It would be interesting to see to
what extent this procedure could be generated to higher orders, both
in kinematics and in the coupling. Assuming that a given correlator
can be expanded in a suitable integral basis, such a procedure could
produce constraints on the expansion coefficients. In the expansion of
four-point correlators around the double-coincidence limit, the leading term will always be
given by the two-magnon operators~\eqref{eq:2magops}, since these are
the states with lowest classical \emph{twisted} dimension among all
operators with non-trivial \emph{anomalous} dimension. Discerning
contributions from different basis integrals at higher loops will
require to include terms at subleading order, which are
provided by internal operators with more than two excitations (and
thus higher twisted dimension).

In order to construct the one-loop OPE, we had to compute
three-point functions of the twisted two-magnon states with
zero-momentum operators. The former are twisted versions of the
two-magnon operators of~\cite{Beisert:2002tn}, while the latter are
twisted versions of R-symmetry descendants of the BMN vacuum
$\trb{Z^J}$. Due to the twisting, the three operators span the whole
scalar $\alg{so}(6)$ sector. It would be interesting to study these
higher-rank three-point functions more thoroughly, and to compare them
with the formulation in terms of hexagon form
factors~\cite{Basso:2015zoa}. In particular, due to the uncommon
representation of the zero-momentum operators under the twisted
conformal symmetry~\eqref{eq:kappa}, correlators involving such
operators have a non-standard spacetime dependence.
More generally, it would be interesting to work out the implications
for correlation functions (Ward identities) of operators that
transform non-trivially under the twisted special conformal generator $\hat\kappa_\mu$.

It would also be interesting to study the correlators of twisted operators
at strong coupling. Using embedding coordinates $Y^I$, $I=1,\dots,6$
for $\grp{AdS}_5$, the boundary Minkowski space, parametrized by
$x^\mu$, is mapped to the lightcone
\begin{equation}
Y=\brk{x^\mu,\half\brk{1+x^2},\half\brk{1-x^2}}
\,.
\end{equation}
This shows that for the vacuum operators $\trb{\phi^J}$, the internal
coordinate $X$~\eqref{eq:Xphi} is identified with the spacetime coordinate $Y$
(up to factors of $i$ due to the different signatures, and the overall
normalization). This is a direct consequence of the definition via the
twisted translation~\eqref{eq:phibytrans}. In the dual sigma model on
$\grp{AdS}_5\times\grp{S}^5$, setting $X=Y$ amounts to equating the
coordinates on the sphere with the AdS coordinates. It would be
interesting to study solutions of this ``reduced'' sigma model.
Perhaps classical solutions corresponding to three-point or
higher-point correlators can be found.

\medskip
\noindent
While this work was being completed, two very interesting
papers~\cite{Eden:2016xvg,Fleury:2016ykk} proposed a decomposition of
twisted correlators (exactly of the type studied here) in terms of the
hexagon form factors constructed earlier in the context of three-point
functions~\cite{Basso:2015zoa}. In particular, the
work~\cite{Fleury:2016ykk} provides a systematic description of
four-point functions (including the spacetime dependence) in terms of
integrability data, without reference to the OPE. Interestingly, Fleury and
Komatsu~\cite{Fleury:2016ykk} observe that they need to exclude
certain ``one-edge reducible'' graphs from their sum over propagator
structures in order to reproduce the correct perturbative results. It
appears plausible that this observation is related to our finding that
certain extremal and all double-trace operators cancel out
in the OPE sum.

Finally, let us end on a note about the fate of $\superN=4$ sYM
integrability beyond the planar limit. Higher-point correlation
functions are inherently suppressed by powers of $1/\Nc$ compared to
two-point functions: Counting powers of $1/\Nc$, four-point functions
on the sphere are of the same order as two-point functions on the
torus. In fact, the two are related by a complete state sum on the
side of the four-point function. If indeed planar four-point functions
can be computed with integrability, tackling non-planar processes by
integrability techniques may finally come within reach.

\subsubsection*{Acknowledgments}

I thank Niklas Beisert and Juan Maldacena for several illuminating
discussions, and the IAS Princeton for support during the initial stage
of this work. This work was in part supported by a Marie
Curie International Outgoing Fellowship within the 7$^{\mathrm{th}}$
European Community Framework Programme under Grant
No.~PIOF-GA-2011-299865.

\appendix

\section{Details on Tensors}
\label{sec:details-tensors}

\paragraph{Useful Identities.}

\begin{equation}
I_{\alpha\gamma}(x)I^\gamma{}_\beta(x)=\eta_{\alpha\beta}
\end{equation}
\begin{equation}
Y_\mu(x_{ij},x_{ik})Y^\mu(x_{ij},x_{ik})=\frac{x_{jk}^2}{x_{ij}^2x_{ik}^2}
\end{equation}
\begin{equation}
I_\rho{}^\sigma(x_{ij})Y_\sigma(x_{ij},x_{ik})=\frac{x_{jk}^2}{x_{ik}^2}Y_\rho(x_{ji},x_{jk})
\end{equation}
%

\paragraph{Coefficients of the Universal Prefactor.}

Using the shorthand notation
\begin{equation}
\check x = \frac{x}{x^2}
\,,
\end{equation}
the coefficient $\cR^{1,2}$~\eqref{eq:R12} reads
\begin{multline}
\cR^{1,2,\mu\nu}
=
4\check x_{13}^\mu \check x_{24}^\nu
-4s\check x_{12}^\mu \check x_{12}^\nu
+4t\check x_{14}^\mu \check x_{23}^\nu
+2(-1+s+t)\lrbrk{
 \check x_{12}^\nu \check x_{14}^\mu
-\check x_{12}^\mu \check x_{23}^\nu
}
\\
+2(1+s-t)\lrbrk{
 \check x_{12}^\nu \check x_{13}^\mu
-\check x_{12}^\mu \check x_{24}^\nu
}
+2(-1+s-t)
\lrbrk{
 \check x_{13}^\mu \check x_{23}^\nu
+\check x_{14}^\mu \check x_{24}^\nu
}
\,.
\label{eq:R12check}
\end{multline}
Similarly,
\begin{multline}
\cR^{1,4,\mu\sigma}
=
-4\check x_{13}^\mu \check x_{24}^\sigma
-4s\check x_{12}^\mu \check x_{34}^\sigma
-4t\check x_{14}^\mu \check x_{14}^\sigma
+2(-1+s+t)\lrbrk{
 \check x_{12}^\mu \check x_{14}^\sigma
+\check x_{14}^\mu \check x_{34}^\sigma
}
\\
+2(1+s-t)\lrbrk{
 \check x_{12}^\mu \check x_{24}^\sigma
+\check x_{13}^\mu \check x_{34}^\sigma
}
-2(-1+s-t)\lrbrk{
 \check x_{13}^\mu \check x_{14}^\sigma
+\check x_{14}^\mu \check x_{24}^\sigma
}
\,.
\label{eq:R14}
\end{multline}
In fact, all $\cR^{i,j}$ can be obtained
from $\cR^{1,2}$ by applying the following replacements:
\begin{align}
\cR^{1,3,\mu\rho} :&\quad
\check x_{12}^\nu\to -\check x_{34}^\rho,
\check x_{23}^\nu\to -\check x_{23}^\rho,
\check x_{24}^\nu\to -\check x_{13}^\rho
\\
\cR^{1,4,\mu\sigma} :&\quad
\check x_{12}^\nu\to +\check x_{34}^\sigma,
\check x_{23}^\nu\to -\check x_{14}^\sigma,
\check x_{24}^\nu\to -\check x_{24}^\sigma
\\
\cR^{2,3,\nu\rho} :&\quad
\check x_{12}^\mu\to +\check x_{34}^\rho,
\check x_{13}^\mu\to -\check x_{13}^\rho,
\check x_{14}^\mu\to -\check x_{23}^\rho
\\
\cR^{2,4,\nu\sigma} :&\quad
\check x_{12}^\mu\to -\check x_{34}^\sigma,
\check x_{13}^\mu\to -\check x_{24}^\sigma,
\check x_{14}^\mu\to -\check x_{14}^\sigma
\\
\cR^{3,4,\rho\sigma} :&\quad
\text{combinations of replacements for $\cR^{1,3,\mu\rho}$ and $\cR^{2,4,\nu\sigma}$}
\nn\\
&\quad \text{(or combinations of replacements for $\cR^{1,4,\mu\sigma}$ and $\cR^{2,3,\nu\rho}$.)}
\end{align}
In the double-coincidence limit
$\abs{x_{12}}\abs{x_{34}}\ll\abs{x_{23}}$, the coefficients
$\cR^{i,j}$ become
\begin{align}
\cR^{1,2}_{\mu\nu}
&=
-2\bigbrk{
I_{12,\mu\gamma}I_{23}{}^\gamma{}_{\alpha}I_{23,\nu\beta}
+x_{12,\mu}\check x_{12,\nu}\eta_{\alpha\beta}
+(\mu\leftrightarrow\nu)
}
x_{23}^{-4}x_{34}^\alpha x_{34}^\beta
\,,
\label{eq:R12limit}
\\
\cR^{3,4}_{\mu\nu}
&=
-2\bigbrk{
I_{34,\mu\gamma}I_{23}{}^\gamma{}_{\alpha}I_{23,\nu\beta}
+x_{34,\mu}\check x_{34,\nu}\eta_{\alpha\beta}
+(\mu\leftrightarrow\nu)
}
x_{23}^{-4}x_{12}^\alpha x_{12}^\beta
\,,
\\
\cR^{1,3}_{\mu\rho}
&=
4x_{12}^\alpha
\Bigbrk{
\half I_{23,\alpha\beta}\bigbrk{I_{12,\mu\gamma}I_{23}{}^\gamma{}_{\rho}+I_{23,\mu\gamma}I_{34}{}^\gamma{}_{\rho}-2I_{23,\mu\rho}}
+I_{23,\mu\beta}I_{23,\rho\alpha}
+\eta_{\mu\alpha}\eta_{\rho\beta}
}
x_{23}^{-4}x_{34}^\beta
\,,
\\
\cR^{1,4}_{\mu\sigma}
&=
\cR^{2,3}_{\mu\sigma}
=
-\cR^{2,4}_{\mu\sigma}
=
-\cR^{1,3}_{\mu\sigma}
\,.
\label{eq:R14limit}
\end{align}
Next, the coefficients with three indices:
\begin{multline}
\cR^{1,2,3,\mu\nu\rho}
=
\\
\sqrt{2} \Bigsbrk{
2 t \check x_{14}^\mu \bigbrk{-\eta^{\rho\nu}/x_{23}^2 - 4\check x_{23}^\rho\check x_{23}^\nu}
-2 \check x_{24}^\nu \bigbrk{\eta^{\rho\mu}/x_{13}^2 + 4\check x_{13}^\rho\check x_{13}^\mu}
+2 s \check x_{34}^\rho \bigbrk{-\eta^{\mu\nu}/x_{12}^2 - 4\check x_{12}^\mu\check x_{12}^\nu}
\\
+(-1 + s + t) \bigbrk{
\eta^{\rho\nu}\check x_{12}^\mu/x_{23}^2
-\eta^{\mu\nu}\check x_{23}^\rho/x_{12}^2
-2\check x_{12}^\nu\check x_{14}^\mu Y^\rho_{23,34}
-2\check x_{23}^\nu\check x_{34}^\rho Y^\mu_{12,41}
}
\\
+(1 + s - t) \bigbrk{
-\eta^{\rho\mu}\check x_{12}^\nu/x_{13}^2
-\eta^{\mu\nu}\check x_{13}^\rho/x_{12}^2
+2\check x_{12}^\mu\check x_{24}^\nu Y^\rho_{13,34}
+2\check x_{13}^\mu\check x_{34}^\rho Y^\nu_{12,24}
}
\\
+(-1 + s - t) \bigbrk{
-\eta^{\rho\nu}\check x_{13}^\mu/x_{23}^2
-\eta^{\rho\mu}\check x_{23}^\nu/x_{13}^2
-2\check x_{23}^\rho\check x_{24}^\nu Y^\mu_{13,41}
-2\check x_{13}^\rho\check x_{14}^\mu Y^\nu_{23,42}
}
}
\end{multline}
The coefficient $\cR^{1,2,4,\mu\nu\rho}$ is obtained from $\cR^{1,2,3,\nu\mu\rho}$
(note the permuted indices) by the replacements:
\begin{equation}
\check x_{13} \leftrightarrow \check x_{24}
\,,\quad
\check x_{14} \leftrightarrow \check x_{23}
\,,\quad
x_{34} \to -x_{34}
\,,\quad
x_{12} \to -x_{12}
\,,\quad
\frac{\eta_{\rho\mu}}{x_{13}^2} \to \frac{\eta_{\rho\mu}}{x_{24}^2}
\,,\quad
\frac{\eta_{\rho\nu}}{x_{23}^2} \to \frac{\eta_{\rho\nu}}{x_{14}^2}
\,.
\end{equation}
The coefficient $\cR^{1,3,4,\mu\rho\sigma}$ is obtained from $\cR^{1,2,3,\rho\sigma\mu}$
(note the permuted indices) by the replacements:
\begin{equation}
\check x_{12} \leftrightarrow \check x_{34}
\,,\quad
\check x_{14} \leftrightarrow -\check x_{23}
\,,\quad
x_{24} \to -x_{24}
\,,\quad
x_{13} \to -x_{13}
\,,\quad
\frac{\eta_{\rho\sigma}}{x_{12}^2} \to \frac{\eta_{\rho\sigma}}{x_{34}^2}
\,,\quad
\frac{\eta_{\mu\sigma}}{x_{23}^2} \to \frac{\eta_{\mu\sigma}}{x_{14}^2}
\,.
\end{equation}
The coefficient $\cR^{2,3,4,\nu\rho\sigma}$ is obtained from $\cR^{1,2,3,\sigma\rho\nu}$
(note the permuted indices) by the replacements:
\begin{equation}
\check x_{12} \leftrightarrow -\check x_{34}
\,,\quad
\check x_{13} \leftrightarrow -\check x_{24}
\,,\quad
x_{14} \to -x_{24}
\,,\quad
x_{23} \to -x_{23}
\,,\quad
\frac{\eta_{\nu\sigma}}{x_{13}^2} \to \frac{\eta_{\nu\sigma}}{x_{24}^2}
\,,\quad
\frac{\eta_{\rho\sigma}}{x_{12}^2} \to \frac{\eta_{\rho\sigma}}{x_{34}^2}
\,.
\end{equation}
In the double-coincidence limit
$\abs{x_{12}}\abs{x_{34}}\ll\abs{x_{23}}$, the coefficients
$\cR^{i,j,k}$ become
\begin{align}
\cR^{1,2,3,\mu\nu\rho}
&=
\frac{\sqrt{2}}{x_{23}^4}
\Bigbrk{
-x_{34}^{\rho}
\bigbrk{3\eta^{\mu\nu} - 2 I_{12}^{\mu\nu}}
-2 I_{23}^{\nu\rho}\,I_{23}^\mu\cdot x_{34}
\nn\\
&\mspace{60mu}
+x_{12}\cdot I_{23}\cdot x_{34}
\bigbrk{
\eta^{\mu\nu}\,\check x_{12}\cdot I_{23}^\rho
+2\check x_{12}^\mu
\bigbrk{2\eta^{\rho\gamma} - I_{34}^{\rho\gamma}}
I_{23,\gamma}^\nu
}
}
+(\mu\leftrightarrow\nu)
\,,
\label{eq:R123limit}
\\
\cR^{1,2,4,\mu\nu\rho}
&=
-\cR^{1,2,3,\mu\nu\rho}
\,,
\label{eq:R124limit}
\\
\cR^{1,3,4,\rho\mu\nu}
&=
\frac{\sqrt{2}}{x_{23}^4}
\Bigbrk{
-x_{12}^{\rho}
\bigbrk{3\eta^{\mu\nu} - 2 I_{34}^{\mu\nu}}
-2 I_{23}^{\nu\rho}\,I_{23}^\mu\cdot x_{12}
\nn\\
&\mspace{60mu}
+x_{12}\cdot I_{23}\cdot x_{34}
\bigbrk{
\eta^{\mu\nu}\,\check x_{34}\cdot I_{23}^\rho
+2\check x_{34}^\mu
\bigbrk{2\eta^{\rho\gamma} - I_{12}^{\rho\gamma}}
I_{23,\gamma}^\nu
}
}
+(\mu\leftrightarrow\nu)
\,,
\label{eq:R134limit}
\\
\cR^{2,3,4,\rho\mu\nu}
&=
-\cR^{1,3,4,\rho\mu\nu}
\,.
\label{eq:R234limit}
\end{align}
Finally,
\begin{align}
\cR^{1,2,3,4,\mu\nu\rho\sigma}&=
16 \check x_{13}^\rho \check x_{13}^\mu \check x_{24}^\sigma \check x_{24}^\nu
+ 8 \check x_{24}^\sigma \check x_{24}^\nu \eta^{\rho\mu}/x_{13}^2
+ 2 \eta^{\sigma\nu} \bigbrk{4 \check x_{13}^\rho \check x_{13}^\mu + \eta^{\rho\mu}/x_{13}^2}/x_{24}^2
\nn\\
&\mspace{-60mu}
+ \brk{-1 + s - t} \Bigsbrk{
  4 \check x_{13}^\rho \check x_{14}^\mu \check x_{23}^\nu \check x_{24}^\sigma
+ 4 \check x_{13}^\mu \check x_{14}^\sigma \check x_{23}^\rho \check x_{24}^\nu
\nn\\
&
+ 2 \bigbrk{\check x_{13}^\rho \check x_{23}^\nu + \check x_{13}^\rho \check x_{24}^\nu + \check x_{23}^\rho \check x_{24}^\nu} \eta^{\sigma\mu}/x_{14}^2
+ 2 \bigbrk{\check x_{13}^\rho \check x_{14}^\mu + \check x_{13}^\mu \check x_{23}^\rho + \check x_{14}^\mu \check x_{23}^\rho} \eta^{\sigma\nu}/x_{24}^2
\nn\\
&
+ 2 \bigbrk{\check x_{14}^\sigma \check x_{23}^\nu + \check x_{23}^\nu \check x_{24}^\sigma + \check x_{14}^\sigma \check x_{24}^\nu} \eta^{\rho\mu}/x_{13}^2
+ 2 \bigbrk{\check x_{13}^\mu \check x_{14}^\sigma + \check x_{13}^\mu \check x_{24}^\sigma + \check x_{14}^\mu \check x_{24}^\sigma} \eta^{\rho\nu}/x_{23}^2
}
\nn\\
&\mspace{-60mu}
+ \brk{-1 + s + t} \Bigsbrk{
  4 \check x_{12}^\nu \check x_{14}^\mu \check x_{23}^\rho \check x_{34}^\sigma
+ 4 \check x_{12}^\mu \check x_{14}^\sigma \check x_{23}^\nu \check x_{34}^\rho
\nn\\
&
+ 2 \bigbrk{-\check x_{12}^\nu \check x_{14}^\mu + \check x_{12}^\mu \check x_{23}^\nu + \check x_{14}^\mu \check x_{23}^\nu} \eta^{\sigma\rho}/x_{34}^2
+ 2 \bigbrk{\check x_{12}^\nu \check x_{23}^\rho - \check x_{12}^\nu \check x_{34}^\rho + \check x_{23}^\nu \check x_{34}^\rho} \eta^{\sigma\mu}/x_{14}^2
\nn\\
&
+ 2 \bigbrk{-\check x_{12}^\mu \check x_{14}^\sigma - \check x_{12}^\mu \check x_{34}^\sigma - \check x_{14}^\mu \check x_{34}^\sigma} \eta^{\rho\nu}/x_{23}^2
+ 2 \bigbrk{\check x_{14}^\sigma \check x_{23}^\rho + \check x_{23}^\rho \check x_{34}^\sigma - \check x_{14}^\sigma \check x_{34}^\rho} \eta^{\mu\nu}/x_{12}^2
}
\nn\\
&\mspace{-60mu}
+ \brk{1 + s - t} \Bigsbrk{
- 4 \check x_{12}^\mu \check x_{13}^\rho \check x_{24}^\nu \check x_{34}^\sigma
- 4 \check x_{12}^\nu \check x_{13}^\mu \check x_{24}^\sigma \check x_{34}^\rho
\nn\\
&
+ 2 \bigbrk{-\check x_{12}^\nu \check x_{13}^\mu + \check x_{12}^\mu \check x_{24}^\nu + \check x_{13}^\mu \check x_{24}^\nu} \eta^{\sigma\rho}/x_{34}^2
+ 2 \bigbrk{-\check x_{12}^\mu \check x_{13}^\rho + \check x_{12}^\mu \check x_{34}^\rho + \check x_{13}^\mu \check x_{34}^\rho} \eta^{\sigma\nu}/x_{24}^2
\nn\\
&
+ 2 \bigbrk{\check x_{12}^\nu \check x_{24}^\sigma + \check x_{12}^\nu \check x_{34}^\sigma - \check x_{24}^\nu \check x_{34}^\sigma} \eta^{\rho\mu}/x_{13}^2
+ 2 \bigbrk{\check x_{13}^\rho \check x_{24}^\sigma + \check x_{13}^\rho \check x_{34}^\sigma - \check x_{24}^\sigma \check x_{34}^\rho} \eta^{\mu\nu}/x_{12}^2
}
\nn\\
&\mspace{-60mu}
+ s \Bigsbrk{
16 \check x_{12}^\mu \check x_{12}^\nu \check x_{34}^\sigma \check x_{34}^\rho
+ 8 \check x_{34}^\sigma \check x_{34}^\rho \eta^{\mu\nu}/x_{12}^2
+ 2 \eta^{\sigma\rho} \bigbrk{4 \check x_{12}^\mu \check x_{12}^\nu + \eta^{\mu\nu}/x_{12}^2}/x_{34}^2
}
\nn\\
&\mspace{-60mu}
+ t \Bigsbrk{
  16 \check x_{14}^\sigma \check x_{14}^\mu \check x_{23}^\rho \check x_{23}^\nu
+ 8 \check x_{14}^\sigma \check x_{14}^\mu \eta^{\rho\nu}/x_{23}^2
+ 2 \eta^{\sigma\mu} \bigbrk{4 \check x_{23}^\rho \check x_{23}^\nu + \eta^{\rho\nu}/x_{23}^2}/x_{14}^2
}
\,.
\end{align}
In the double-coincidence limit, this becomes
\begin{multline}
\cR^{1,2,3,4,\mu\nu\rho\sigma}=
\frac{1}{x_{23}^4}
\Bigbrk{
-2\,x_{12}\cdot I_{23}\cdot x_{34}
\bigbrk{
\check x_{12}^\mu\,I_{23}^\nu\cdot\check x_{34}\,\eta^{\rho\sigma}
+\eta^{\mu\nu}\check x_{12}\cdot I_{23}^\rho\,\check x_{34}^\sigma
+2\check x_{12}^\mu I_{23}^{\nu\rho}\check x_{34}^\sigma
}
\\
+I_{23}^{\mu\rho}I_{23}^{\nu\sigma}
+\bigbrk{I_{12}^{\mu\nu}-2\eta^{\mu\nu}}\bigbrk{I_{34}^{\rho\sigma}-2\eta^{\rho\sigma}}
-\half\eta^{\mu\nu}\eta^{\rho\sigma}
}
+(\mu\leftrightarrow\nu)
+(\rho\leftrightarrow\sigma)
\end{multline}
%

\pdfbookmark[1]{\refname}{references}
\bibliographystyle{nb}
\bibliography{twistcorr}

\end{document}